\documentclass[aps,prx,twocolumn,amsmath,amssymb,floatfix,superscriptaddress]{revtex4-1}
\usepackage{tabularx}
\usepackage{bm}
\usepackage{euscript}
\usepackage{epsfig,psfrag,subfigure}
\usepackage{graphicx}
\usepackage{color}
\usepackage{amsfonts}
\usepackage{exscale}
\usepackage{amsbsy}
\usepackage{multirow}

\def\avg#1{\left\langle#1\right\rangle}

\def\be{\begin{equation}}       \def\ee{\end{equation}}
\def\bea{\begin{eqnarray}}      \def\eea{\end{eqnarray}}
\def\ba{\begin{array} }
\def\ea{\end{array} }

\def\nn{\nonumber}

\def\=>{\Rightarrow}
\def\>{\rightarrow}
\def\A{\uparrow}
\def\V{\downarrow}

\def\Fig#1{Fig.~\ref{#1}}

\renewcommand{\>}{\rangle}

\begin{document}

\title{Ground state phase diagram of the doped Hubbard model on the 4-leg cylinder}

\author{Yi-Fan Jiang}
\affiliation{Stanford Institute for Materials and Energy Sciences, SLAC National Accelerator Laboratory and Stanford University, Menlo Park, CA 94025, USA}

\author{Jan Zaanen}
\affiliation{Instituut-Lorentz for Theoretical Physics , Universiteit Leiden, P.O. Box 9506, 2300 RA Leiden, The Netherlands}
\affiliation{Department of Physics, Stanford University, Stanford CA 94305, USA}

\author{Thomas P. Devereaux}
\affiliation{Stanford Institute for Materials and Energy Sciences, SLAC National Accelerator Laboratory and Stanford University, Menlo Park, CA 94025, USA}
\affiliation{{Department of Materials Science and Engineering, Stanford University, Stanford, CA 94305, USA}}

\author{Hong-Chen Jiang}
\affiliation{Stanford Institute for Materials and Energy Sciences, SLAC National Accelerator Laboratory and Stanford University, Menlo Park, CA 94025, USA}

\begin{abstract}
We study the ground state properties of the  Hubbard model on a 4-leg cylinder with doped hole concentration per site $\delta\leq 12.5\%$ using density-matrix renormalization group. By keeping a large number of states for long system sizes, we find that the nature of the ground state is remarkably sensitive to the presence of next-nearest-neighbor hopping $t'$. 
Without $t'$ the ground state of the system corresponds with the insulating filled stripe phase with long-range charge-density-wave (CDW) order and short-range incommensurate spin correlations appears. However, for a small negative $t'$ a phase characterized by coexisting algebraic d-wave superconducting (SC)- and algebraic CDW correlations. In addition, it shows short range spin- and fermion correlations consistent with a canonical Luther-Emery (LE) liquid, except that the charge- and spin periodicities are consistent with half-filled stripes instead of the $4 k_F$ and $2 k_F$ wavevectors generic for one dimensional chains. For a small positive $t'$ yet another phase takes over showing similar SC and CDW correlations. However, the fermions are now characterized by a (near) infinite correlation length while the gapped spin system is characterized  by simple staggered antiferromagnetic correlations. We will show that this is consistent with a LE formed from a weakly coupled (BCS like) d-wave superconductor on the ladder where the interactions have only the effect to stabilize a cuprate style magnetic resonance. 
 \end{abstract}
\date{\today}

\maketitle

\section {Introduction}

The Hubbard model is the simplest model that captures essential features of strongly interacting electrons realized in solids containing transition metal and/or rare earth elements, characterized by a compromise between kinetic energy and strong local repulsion. It describes a tight binding model for the band structure with hopping matrix elements $t_{ij}$, supplemented by an on-site Coulomb repulsion $U$,
\begin{equation}
H = \sum_{ij, \sigma} t_{ij} c^{\dagger}_{i \sigma} c_{j \sigma} + U \sum_i n_{i \uparrow} n_{i \downarrow},\label{Eq:Ham}
\end{equation}
where $c^\dagger_{i\sigma}$ is the electron creation operator on site $i=(x_i,y_i)$ with spin $\sigma$, $n_i=\sum_{\sigma} c_{i\sigma}^\dagger c_{i\sigma}$ is the electron number operator. In this paper, we will consider a square lattice geometry compactified on a cylinder, specializing the hopping integral to $t_{ij}=t$ for nearest-neighbor (NN) and $t_{ij}=t^\prime$ for next-nearest-neighbor (NNN) sites, respectively.

This is among the most studied models in the history of physics in the case of fermions. Its simple appearance is delusional; after tens of thousands of papers it has been brought under mathematical control in one- and infinite dimensions.\cite{Dagotto1994,Lee2006,Fradkin2012,Fradkin2015,Keimer2015,Robinson2019} However, due to spectacular progress in the computational methodology recently some solid results were reported \cite{Blankenbecler1981,White1997,Tocchio2008,Yang2009,Chou2010,Gull2012,Gull2013,Corboz2014,LeBlanc2015,Ehlers2017,Zheng2017,Jiang2018hub,Jiang2018tj,Huang2017,Huang2018,Darmawan2018}. In part this is due to the exponential growth of computational resources having the effect that for instance Quantum Monte Carlo (QMC) now yields results at temperatures low enough to be of physical relevance.\cite{Huang2017,Huang2018} On the other hand, new algorithms were developed inspired from quantum information where the difficulty with, e.g. the Hubbard model is to keep track of the many-body entanglement.
The density-matrix renormalization group (DMRG) is the oldest of this family.\cite{White1992} While it was designed to study systems in one dimension (1D), it has been made to work well for quasi-1D systems having a finite spatial extend in the transverse direction, i.e., the ladder systems, which is now the standard setting to compute the physics in two dimensions (2D).\cite{Stoudenmire2012}

A key ingredient of DMRG is the degree of many-body quantum entanglement which is controlled by the bond dimension, i.e., the number of states $m$ kept in the local matrices.\cite{White1992,Stoudenmire2012} It turns out that for the Hubbard model on 4-leg cylinders the physics is quite sensitive to this many-body entanglement: bond dimensions as large as $m$=20000 are required to capture correct long-distance physics and converged ground states.\cite{Ehlers2017,Zheng2017,Jiang2018hub,Jiang2018tj} A holy grail is whether superconductivity occurs in the Hubbard model. There is no sign of it at small bond dimensions but it emerges when  this becomes large enough. We reported recently on the presence of a Luther-Emery (LE) liquid characterized by algebraic superconducting (SC) order coexisting with the (dual) algebraic charge-density-wave (CDW) order in a particular parameter regime of both the Hubbard- and closely related $t$-$J$ model where the critical role of $t'$ was shown to control the pattern of stripes, and as a consequence, superconductivity.\cite{Jiang2018hub,Jiang2018tj} Here we will report on how the ground states of the Hubbard model look like in a much larger regime of physically relevant parameters.\cite{Hirayama2018,Hirayama2019} 

Another aspect is that already the very first 4-leg cylinder results showed a strong appetite to form ``intertwined order'', specifically of the spin-stripe variety. \cite{Dodaro2017,Huang2017,Huang2018,Jiang2018hub,Jiang2018tj} The stripes have a long history going back to the 1980's when it was discovered that according to classic mean-field theory (Hartree-Fock) in the case of large $U$ the doped systems develop rather complex textures: domain  walls form in the antiferromagnetic spin background, where the carriers localize forming ``rivers of charge''.\cite{Zaanen1989}  According to the mean-field theory, these stripes are stable under the condition that there is one carrier per domain-wall unit cell, for the reason that a gap opens at $E_F$ rendering these stripe phases to be insulating. The outcomes of various methods characterized by qualitatively different systematic errors were compared and such insulating filled stripes ground states were established. This includes the DMRG results on the 4-leg cylinders at $t'=0$ and $U=8t$.\cite{LeBlanc2015,Ehlers2017,Zheng2017,Jiang2018hub}

All along there were indications that various ground states of a quite different nature may be quite close in energy in Hubbard and $t$-$J$ models.\cite{White1997,Corboz2014,Zheng2017} The $t$-$J$ model has the advantage that it can be easily deformed involving longer range exchange interactions and so forth. As a function of these extra parameters a wealth of different phases, ``stripy'' and otherwise,  were found on the 4-leg cylinder.\cite{White1997,Scalapino2012,Jiang2018hub,Jiang2018tj} The early ladder results showed also such stripes with the difference that these turned out to be ``half-filled'': one carrier is associated with {\em two} domain wall unit cells.\cite{White1997,White1999,Scalapino2012} Meanwhile, such stripe instabilities were observed in high-Tc superconductors of the so-called 214 family.\cite{Fradkin2015} Strikingly, these are also half-filled giving further support for the approach in general. 

Using the large bond dimension DMRG simulations we will report here a surprise. The nature of the ground states of the doped Hubbard system appears to be {\em extremely sensitive
to the next nearest neighbour hopping $t'$}. We can diagnose the various phases in a precise fashion, the reason being that these are 1D systems that will reveal 
the universal behaviours of such systems at long length scales. However, different from the canonical chain systems at short distances the ladders are more like 2D systems.
This will alter the ``numbers'' at long distances which in turn may reveal aspects of relevance to the 2D case, given that the convergence as function of ladder width is expected to be 
rather rapid. 

\begin{figure}[bt]
\centering
\includegraphics[width=\linewidth]{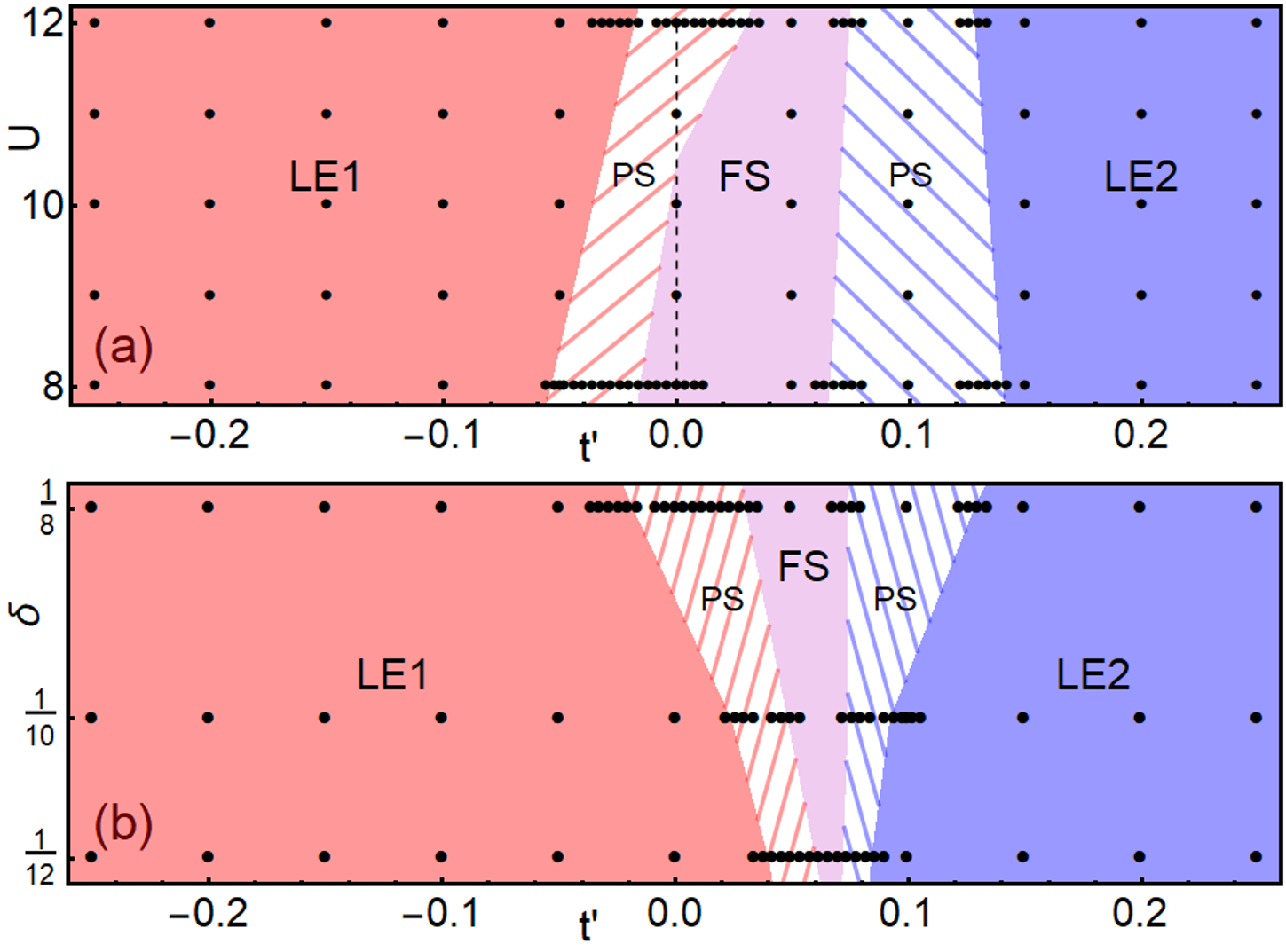}
\caption{(Color online) Ground state phase diagram of the Hubbard model in Eq.(\ref{Eq:Ham}) in (a) as a function of $U$ and $t'$ at hole doping concentration $\delta=12.5\%$ where the dashed line labels $t'=0$, and in (b) as a function of $\delta$ and $t'$ at $U=12$. Here $t=1$ and the black dots are the data points.}
\label{Fig:PhaseDiagram}
\end{figure}

\begin{figure}[bt]
\centering
\includegraphics[width=\linewidth]{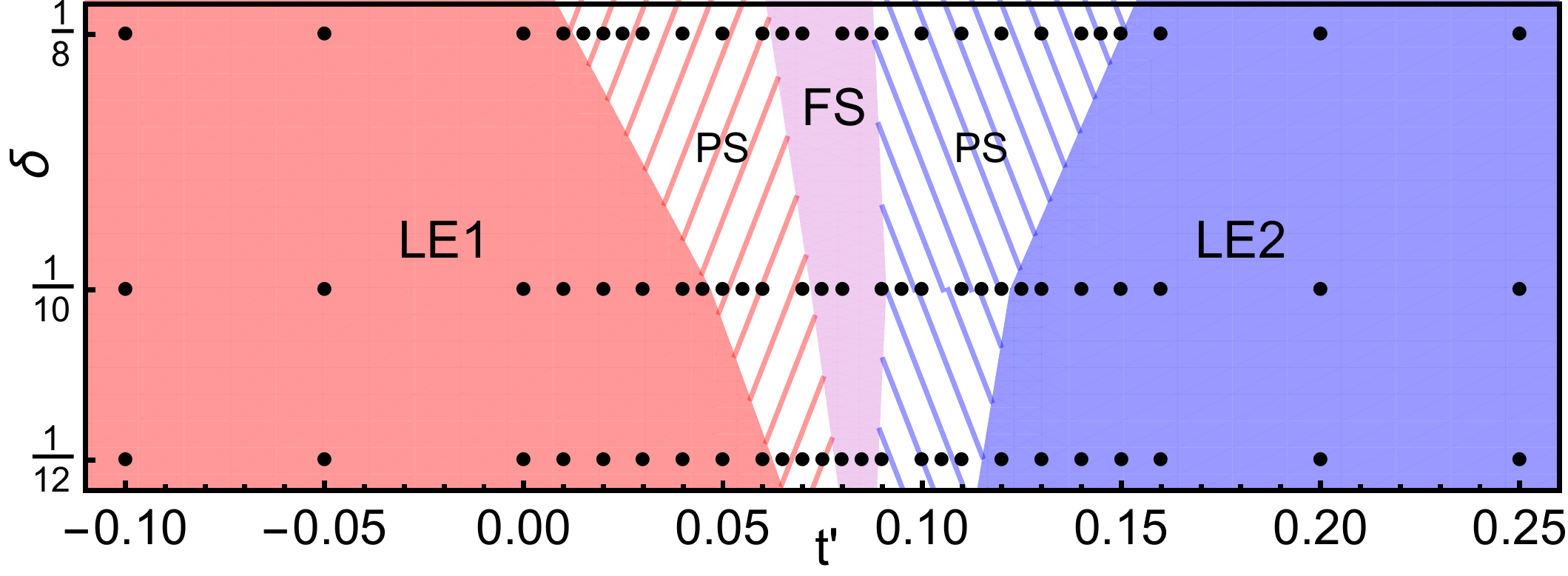}
\caption{(Color online) Ground state phase diagram of the $t-t'-J$ model in Eq.(\ref{Eq:HamtJ}) as a function of hole doping concentration $\delta$ and $t'$. The black dots are the data points. Here $t=1$ and $J=1/3$.}
\label{Fig:PhaseDiagramtJ}
\end{figure}

The main result is a zero-temperature phase diagram which is surprisingly sensitive to a single parameter: $t'$, Fig.\ref{Fig:PhaseDiagram}. Not much happens upon varying the doping $\delta$ and $U$ quite a bit: it looks similar for the very strong coupling $t-t'-J$ model in Fig.\ref{Fig:PhaseDiagramtJ}. However, varying $t'$ by a tenth of $t$ has the effect of stabilizing completely different ground states. Using the 1D diagnostics, the ``purple'' regime for $t' \simeq 0$ is identified as the commensurately pinned ``Luttinger liquid'' charge density/spin density wave that coincides with the filled stripes \cite{Zaanen1989} in the 2D language. The phases found at positive- and negative $t'$ are in the first place characterized by coexisting algebraic SC- and CDW order which are in dual relation: {\em the} signature of the LE phase, and we call these therefore the LE1 and LE2 phases, indications for which were already reported in \cite{Jiang2018hub} and \cite{Dodaro2017,Huang2018}. 

In other regards the LE1 and LE2 phases are however entirely different. The LE1 phase as first reported in  \cite{Jiang2018hub} is quite like the canonical LE phase realized on chains. The spin correlations are short ranged, associated with spin gap characterizing the Cooper pairs while they show a periodicity which is twice the charge periodicity. However, the latter counts like half-filled stripes and it is no longer related to the $2k_F$ and $4k_F$ spin- and charge periodicities hardwired in the chain systems. %However, the $LE2$ appears to violate these principles. 
The equal-time fermion propagator demonstrates that the gap in their spectrum is very small, if not even vanishing. In addition, the spin correlations are short ranged but characterized by a simple two sublattice periodicity!

We will argue that this can be naturally explained asserting that this LE2 phase is a smooth continuation from a weakly coupled (BCS) $d$-wave superconductor living on the ladder. 
It is easy to see that the vanishing fermion mass revealed by the equal time fermion correlator is just anchored in the fact that the electronic modes of the ladder will touch the nodal direction where the d-wave gap is vanishing. Turning to the spins, it is well established that in weakly interacting 1D systems Random Phase Approximation (RPA) is quite accurate. According to RPA the massless spin fluctuations carry so little spectral weight that these are not seen in the equal time correlations. Instead, it is well established principle that spin fluctuations in a conventional $d$-wave superconductor are susceptible to strong enhancements at staggered wave vectors: the prevalent explanation for the magnetic resonance in near-optimal cuprate superconductors. In the conclusions we will further discuss potential relations between these findings and the situation in these experimental systems.

%==Fig.1 Phase diagram of Hubbard model==

%==Fig.2 Phase diagram of t-t'-J model==

%==Model and Method==
\section{The Phase diagram: overview.}

{\em The methodology.} The DMRG method \cite{White1992} has acquired quite a merit for the computation of the ground states of strictly 1D (chain) systems. Although the 
computations become rapidly more demanding, it was early on realized that it can be mobilized to systems with a transversal extent and the results converge relatively rapidly towards the 2D physics.\cite{Stoudenmire2012} Two-leg ladders are clearly too narrow while it was very recently found that one has to employ very large bond dimensions to get fully convergent results on four-leg ladders. Here we employ such ladders of width  (number of sites) $L_y =4$ with cylindrical boundary condition in this compact direction, and open boundary condition in the extended direction with a length up to $L_x=96$. We keep up to $m=20000$ states in each DMRG block with a typical truncation error $\epsilon\sim 10^{-6}$ and perform around 60 sweeps, which leads to excellent convergence of our results. Further details are provided in the Supplemental Material (SM).

%A limitation of DMRG is that only the ground state is computed. %As a ramification, 
While we have access only to {\em equal time} correlation functions %.However, 
it is a matter of principle that at long distances this system has to behave like a 1D system. These are exceptional in the regard that their IR fixed points are well understood.\cite{Giamarchi2004} Information on their equal time two point correlation functions suffices to reconstruct the nature of these fixed points: the combination of the electron density, spin, pair and single electron two points functions augmented by the bipartite entanglement entropy which we all compute suffices to find out the nature of the ground states.   

As we will highlight, the precise properties of these fixed points may yet be different  from the canonical truly 1D chain systems. The reason is that at short distance 
the finite extent of the ladder in the transversal direction does change the physics qualitatively.  This  short distance (``UV'') physics sets the conditions for the long wavelength properties which is much richer than in 1D systems. At short distance one encounters the vigorous, strongly coupled 2D quantum system. The fact that very large bond dimensions are needed in the DMRG simulations - much larger than truly 1D systems indicates that this is a densely many body entangled affair where the (implicit) semiclassical language underlying the established 1D canon may well fall short. The open question is whether this will be eventually understood in terms of analytical theory and/or general emergence principle.

{\em The models.} Our main focus is on the Hubbard model Eq. (\ref{Eq:Ham}). The main interest is in the regime of low doping relative to the Mott insulator realized at
half filling. Given a total number of sites $N=L_x\times L_y$ there is  $n_i=1$ electron per site when $N_e=N$. The hole doping concentration is defined as $\delta=N_h/N$ 
where $N_h=N-N_e$ is the number of doped holes. We focus in on the doping regime $1/12 \le \delta \le 1/8$. In addition, the interest is in the intermediate to large interaction
regime $U > 8t$ where we take all along $t =1$ as unit of energy.  At several instances it will be useful to compare the results for the Hubbard model with those for the $t-t'-J$
model,
\begin{equation}
H = \sum_{ij, \sigma} t_{ij} c^{\dagger}_{i \sigma} c_{j \sigma} + J \sum_{\langle ij\rangle} \vec{S}_i\cdot \vec{S}_j,\label{Eq:HamtJ}
\end{equation}
which captures the low energy  physics of the Hubbard model in the strong coupling regime, $U >> t$. The $c^{\dagger}_{i \sigma} $ creates projected fermions (no double occupancy is
created in the hopping process), while $ \vec{S}_i$ describes Heisenberg spins coupled by a superexchange interaction $J \sim  t^2/U$. 

{\em A remarkable phase diagram.} The main result of our simulation is the zero temperature phase diagram summarized in Fig.'s \ref{Fig:PhaseDiagram}, \ref{Fig:PhaseDiagramtJ}. A first surprise is that it is rather insensitive to the interaction strength (Fig.\ref{Fig:PhaseDiagram}, upper panel). As long as $U$ is larger than the bandwidth it looks very similar. This is further amplified by the fact that it looks similar even for the $t-t'-J$ model  (Fig.\ref{Fig:PhaseDiagramtJ}) which is associated with the large $U$ limit. A next surprise is that at least in the doping range we consider the phase diagram is rather insensitive to the doping  (Fig.\ref{Fig:PhaseDiagram}, lower panel) -- as we will discuss underneath, this is especially quite mysterious with regard to the ``filled stripe'' (FS) phase in the middle. 

The profundity is in the {\em extreme sensitivity} to the next-nearest-neighbour hopping $t'$, the reason to take it as the horizontal axis in the figures. As function of $t'$ we identify three phases with a quite different thermodynamical identity as we will show in detail underneath. The FS phase found for $t' \simeq 0$  can be regarded as the 1D incarnation of an insulator. By varying $t'$ by the tiny amount of $0.1 t$ we find instead LE phases that can be viewed as the 1D versions of SC phases. However, pending the sign of $t'$ these are of a very different nature. Further stressing the fact that these phases are very different is in the fact that our simulations reveal that these phases are separated by phase separation regions (PS, see SM for details). 

Let us now turn to the identification of these three phases, relying on the behaviour of the equal time correlations. In the sequence FS - LE1 - LE2 these 1D phases deviate 
increasingly from the canonical 1D physics associated with chains. Let us discuss therefore these phases in this order. 

%==Fig.3 Filled stripe phase==
\begin{figure}[tb]
\centering
\includegraphics[width=\linewidth]{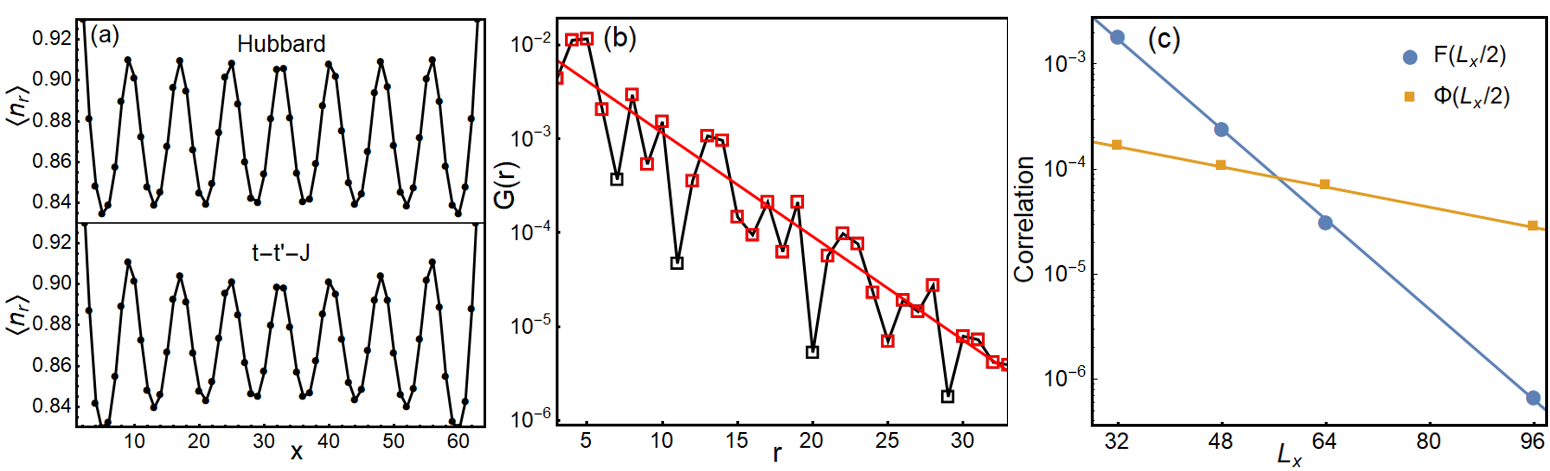}
\caption{(Color online) (a) Charge density profile $n(x)$ of the Hubbard and $t-t'-J$ models in the filled stripe phase. (b) Single particle correlation $G_\sigma(r)$ of the Hubbard model at $U=12$ and $t'=0.05$, whose correlation length $\xi_G\sim 4$ is fitted from the red squares. (c) Superconducting pair-field $\Phi(L_x/2)$ and spin-spin $F(L_x/2)$ correlations of the $t-t'-J$ model. Here $J=1/3$ and $t'=1/12$. The hole doping concentration is $\delta=12.5\%$ for both models.}
\label{n-tp0.05}
\end{figure}

\section{The filled stripe (FS) phase.}

A long time ago it was established that according to Hartree-Fock so-called stripe phases are formed in doped Mott insulators\cite{Zaanen1989,Machida1989,Schulz1990}. 
These consist of antiferromagnetic Mott-insulating domains separated by magnetic domain walls on which the holes localize forming in turn a periodic array. Mean field theory insists that these are most stable at a filling of one hole per domain wall unit cell: the ``filled stripes''. It was quite a surprise when quite recently state of the art numerical calculations (including DMRG) showed these to be the ground state of the Hubbard model at a doping $\delta =1/8$ and $t'=0$.\cite{LeBlanc2015,Ehlers2017,Zheng2017,Jiang2018hub} To give an example, in Fig.\ref{n-tp0.05} we show a result for the Hubbard and $t-t'-J$ model -- these look quite similar everywhere in the purple regime of the phase diagrams  Fig.'s \ref{Fig:PhaseDiagram}, \ref{Fig:PhaseDiagramtJ}. To measure the charge order, we define the local rung density operator as $\hat{n}(x)=\frac{1}{L_y}\sum_{y=1}^{L_y}\hat{n}(x,y)$ 
with expectation value $n(x)=\langle \hat{n}(x)\rangle$. The charge density profile $n(x)$ on a $L_x$=64 cylinder is shown in  Fig.\ref{n-tp0.05}(a), revealing the ordering wave vector $Q_c=2\pi\delta$ consistent with filled stripes.

% The fermion correlation function is defined as 
% \begin{equation}
% G_\sigma(r)=\frac{1}{L_y} \sum_y \avg{c^\dagger_{(x_0,y), \sigma} c_{(x_0+r,y), \sigma}}, \label{Eq:SE}
% \end{equation}
% where $(x_0,y)$ is the reference site and $r$ is the distance between two sites in the $\hat{x}$ direction. The expoential decay of $G_\sigma(r)$ in the filled stripe phase of Hubbard model is shown in Fig.\ref{n-tp0.05}(b), indicating that the system is characterized by a single particle gap. 
%{\bf correct ? You have data? Shouldn't we include fermions in the figure  because they get at center stage in LE1/LE2? I suggest to already introduce the single fermion propagator at this point to include figures for all the cases}
While it still remains challenging to precisely determine the long-distance behavior of the correlations in the Hubbard model even we have kept $m$=20000 states in the DMRG simulation (see SM and \cite{Jiang2018hub} for details), it appears fairly easy to address this issue in the closely related $t-t'-J$ model, where the FS phase is also present.\cite{Dodaro2017} To find out the nature of the SC correlations we study the pair-field correlation, defined as%
 \begin{eqnarray}\label{Eq:SC}
\Phi_{\alpha \beta} (r)=\frac{1}{L_y} \sum_y \avg{\Delta^\dagger_\alpha (x_0,y) \Delta_\beta (x_0+r,y)}, 
\end{eqnarray}
where $\Delta^\dagger_\alpha(x,y) =\frac{1}{\sqrt{2}} \left ( c^\dagger_{(x,y), \A} c^\dagger_{(x,y)+\alpha, \V} - c^\dagger_{(x,y), \V} c^\dagger_{(x,y)+\alpha, \A} \right )$ is the spin-singlet pair-field creation operator and $\alpha= \hat{x}, \hat{y}$ denotes the bond orientations. ($x_0,y$) is the reference bond located at $x_0\sim L_x/4$ and $r$ is the distance between two bonds in $\hat{x}$ direction.
%{\bf Shouldn't we comment on the d-wave nature of the superconductivity ???}
%While it still remains challenging to precisely determine the long-distance behavior of these correlations in the Hubbard model even we have kept around $m$=20000 states in the DMRG simulation (see SM and \cite{Jiang2018hub} for details), it appears fairly easy to address this issue in the closely related $t-t'-J$ model, where the FS phase is also present.\cite{Dodaro2017}. 
We find for a characteristic set of parameters $J=1/3$ and $t'=1/12$ that the pair correlator appears to be short-ranged $\Phi(L_x/2)\sim e^{-L_x/2\xi_{sc}}$ with correlation length $\xi_{sc}\sim 18$ (Fig.\ref{n-tp0.05}) (see SM for more details). These evidences show that the FS phase is an insulator. 
 
We have also calculated the spin-spin correlation%
\begin{eqnarray}
F(r)=\langle \vec{S}_{(x_0,y)}\cdot \vec{S}_{(x_0+r,y)}\rangle \label{Eq:SS}
\end{eqnarray}
where $(x_0,y)$ is the reference site and $r$ is the distance between two sites in the $\hat{x}$ direction. In first instance one would expect 
antiferromagnetism with the incommensurate modulation characterized by $\lambda_s = 2 \lambda_c$ confirming the domain wall picture which is decaying algebraically like a Heisenberg spin chain. However, given the even width of the cylinder this eventually opens up a spin gap of the same kind as on a pure 4-leg Heisenberg ladder. For the Hubbard model\cite{Ehlers2017,Zheng2017,Jiang2018hub}, the spin correlation indeed decays exponentially $F(r)\sim e^{-r/\xi_s}$ with $\xi_s\sim 6.5$.(see SM for detail)  
The one of the $t-t'-J$ model decays in the same way $F(L_x/2)\sim e^{-L_x/2\xi_s}$ as shown in Fig.\ref{n-tp0.05}(c), where we take $r=L_x/2$ for each $L_x$ cylinder for simplicity. 
%{\bf I am puzzled that you claim that this is hard to compute: it should be the vegetable even-rung spin gap that should be relatively large and easy to compute. Surely, the absence of superconductivity is in the present context of way more physical significance.}

Although the literature emphasizes the 2D Hartree-Fock stripes, these ``stripes on the ladder'' may equally well be understood in terms of natural extensions of the strictly 1D theory. These are actually states that are just deformations of the ubiquitous Luttinger liquids. A striking property of strictly 1D systems is that the periodicities seen in charge- and spin correlators are {\em independent} of the interaction strength. In the non-interacting Fermi gas one finds these to be algebraic with ordering wave vectors for spin- and charge set  by the nesting vectors $2k_F$ and $4k_F$, respectively, where $k_F$ is the Fermi wavevector. The only change as function of $U$ is increased is in the exponents governing the algebraic decay.  The reason is revealed by the  $U \rightarrow \infty$ limit, where the notion of the ``squeezed space'' was discovered\cite{Ogata1990}: consider any configuration of holes and spins and remove the sites where the holes are residing and pretend that the spin system living on this ``squeezed'' lattice is described by a Heisenberg model. This squeezing operation can be wired in by the so-called string correlators, and it was shown by DMRG that this works always at long distances, all the way to free limit\cite{Kruis2004}. Very recently this ``squeezed space'' universality was directly verified in cold atom experiments.\cite{Li2016}

The meaning is simple: every charge is a holon, i.e. an electron bound to a kink in the antiferromagnetic spin system. One can view the Luttinger liquid and its decendants (like the LE liquid) as an 1D version of the filled stripes. The simplest way to embed this in 2D is by ``putting holons on a row''.\cite{Zaanen2001} On a ladder geometry this leads automatically to a $2k_F$ spin- and $4k_F$ charge periodicity. The next step is that at dopings like $\delta=1/8$ and $\delta=1/12$ the algebraic CDW order is actually at a higher commensurability with the lattice. In the absence of quantum fluctuations this holon lattice will be subjected to commensurate pinning to the lattice and this pinned crystal is just a ``higher commensurate'' Mott insulator (see \cite{Andrade2018} for a spectacular example revolving around black holes.): the filled stripe phase. 

The fate of an algebraic 1D  solid in a commensurate background potential is enumerated by the Sine-Gordon field theory\cite{Giamarchi2004} that reveals that when the pinning potential compared to the kinetic term enumerating the zero point quantum motions drops below a critical value the pinning seizes to exist with the effect that the electronic density wave depins from the lattice.  This is actually a main difference with the Luther-Emery (LE1,LE2) phases.  Next to being such ``floating charge density waves'' these show automatically also  superconducting correlations. 

The last piece of information comes from the single particle equal time correlator,
\begin{equation}
G_\sigma(r)=\frac{1}{L_y} \sum_y \avg{c^\dagger_{(x_0,y), \sigma} c_{(x_0+r,y), \sigma}}, \label{Eq:SP}
\end{equation}
Consistent with the above interpretation, we find in the Hubbard model its correlation length $\xi_G\sim 4$ to be very short, of order of the width of the ladder. The single particle gap should be larger than the charge commensuration gap combined with the spin gap: given the charge-spin separation principle the electron fractionalizes in a spinon and a holon that can only be inserted at energies larger than the spin- and charge gap respectively. As seen in Fig.\ref{n-tp0.05}(b) it oscillates in a rather complicated, multiharmonic fashion where at least the periodicity of the CDW can be discerned. Given the very small correlation length this may well reflect the rather complex short distance physics and we have not attempted to analyze in further detail.

In summary, the mystery associated with the filled stripes is, why is it so that a minute $t'$ suffices to destabilize it? As Hartree-Fock shows, commensuration is a muscular source of stability \cite{Zaanen1989}. In such a setting the effects of $t'$ would be secondary: it could make a difference when $t'$ becomes of order one, but not for a $t' \sim 0.1$. Apparently the dense entanglement is changing these rules in a way that is beyond our present comprehension.

%==Fig.4==
\begin{figure}[tb]
\centering
\includegraphics[width=\linewidth]{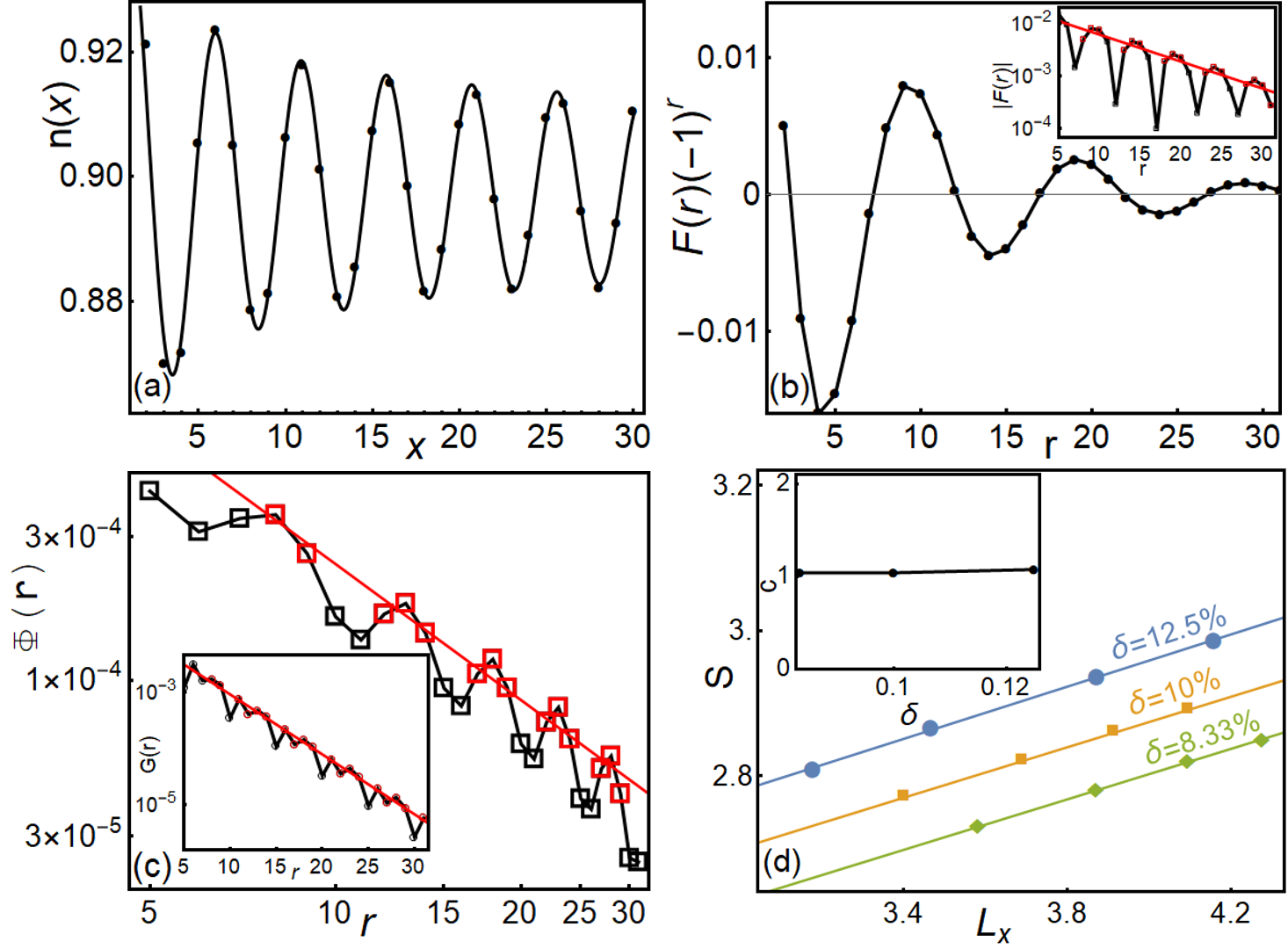}
\caption{(Color online) Ground state properties of the Hubbard model in the LE1 phase at $U=12$, $t'=-0.25$ and $\delta=10\%$, measured on a $L_x=60$ cylinder. (a) Charge density profile $n(x)$ fitted by the Friedel oscillation (solid line) using Eq.(\ref{Eq:CDW}); (b) Spin-spin correlation $F(r)(-1)^r$. Inset is the plot in the semi-logarithmic scale with exponential fitting $|F(r)|\sim e^{-r/\xi_s}$ (red solid line). (c) SC pair-field $\Phi_{yy}(r)$ and single-particle $|G(r)|$ (inset) correlations in the semi-logarithmic scale. Note that only the red data points are used in the fitting. (d) Entanglement entropy $S(L_x/2)$ and the extracted central charge $c$ (inset) at different doping concentration.}
\label{cor-ent-tp-0.25}
\end{figure}

%==Fig.5==
\begin{figure}[tb]
\centering
\includegraphics[width=\linewidth]{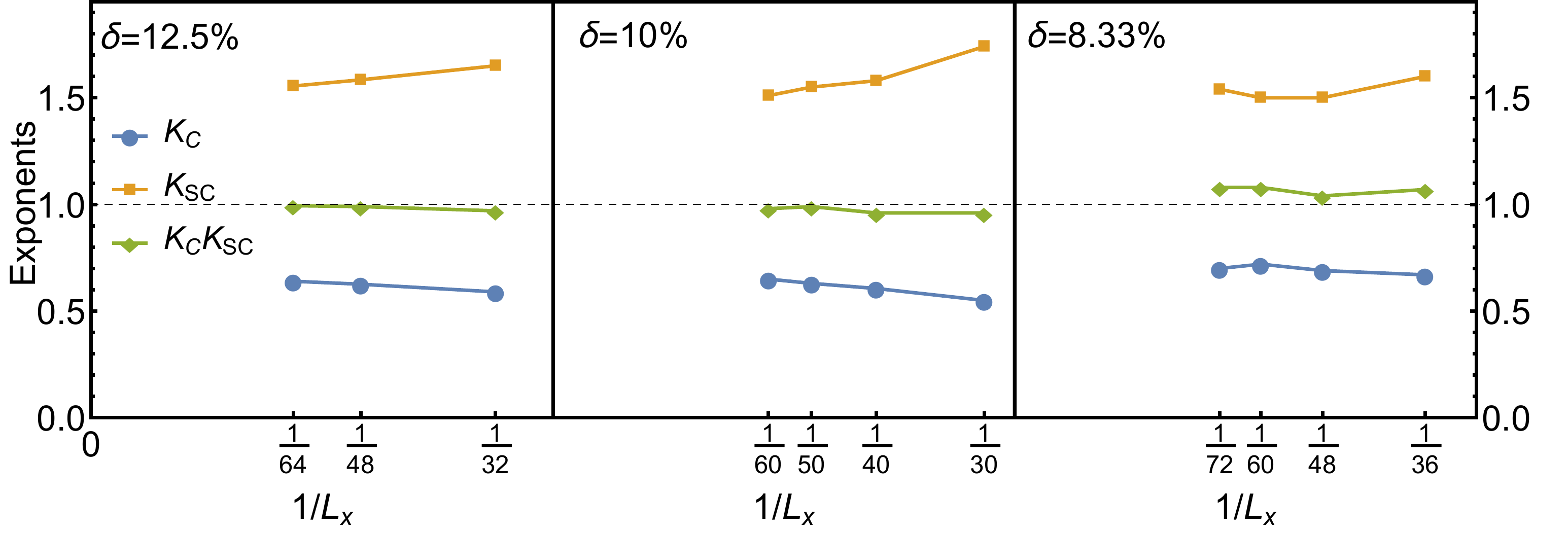}
\caption{(Color online) Exponents $K_c$ and $K_{sc}$ in the LE1 phase of the Hubbard model at various doping concentration $\delta$ as a function of cylinder length $L_x$. Here $t=1$, $U=12$ and $t'=-0.25$.}\label{K-tp-0.25}
\end{figure}

\section{The first Luther-Emery phase (LE1) versus fluctuating stripes.}

The LE1 phase becomes stable already at a very small $t'$, while it appears to be ubiquitous in a large range of $U$'s and dopings only requiring that $t'$ is negative. We already identified it in an earlier study as a posterchild LE phase\cite{Jiang2018hub}.  Luther-Emery  describes the universal SC-like state in 1D\cite{Giamarchi2004}. It arises when attractive interactions are added to a Luttinger liquid. As a signature of the formation
of singlet Cooper pairs a gap opens up both in the single fermion- and spin excitation spectrum. The equal time pair correlation function in Eq.(\ref{Eq:SC}) should show an algebraic behavior $ 1 / x^{K_{sc}}$. However, given that the IR fixed point in 1D is always strongly interacting this goes hand in hand with CDW correlations characterized by a $4k_F$ periodicity
%{\bf Can anybody check? I am not 100\% confident that I remember this correctly, and it is crucially important}  
which also decay algebraically $ 1 / x^{K_c}$. This CDW has to be now in the regime where the lattice commensuration is irrelevant. As a highlight, these exponents should be in a dual relation $K_{sc} = 1 / K_{c}$. The LE1
phase exhibits {\em nearly} all these diagnostic features .

We observe the CDW modulation in the charge density correlator (see Fig.\ref{cor-ent-tp-0.25}(a)). The charge density correlation decays algebraically, where the exponent $K_c$ can be extracted by fitting the Friedel oscillation, which is induced by the open boundaries of the cylinder, of the charge density distribution $n(x)$.\cite{White2002,Jiang2018hub,Jiang2018tj} Specifically, we use%
\begin{eqnarray}
n(x)=n_0 + \delta n\cdot {\rm cos}(2k_F x + \phi)x^{-K_c/2},\label{Eq:CDW}
\end{eqnarray}
to fit the local density profile to extract the Luttinger exponent $K_c$. Here, $\delta n$ is the non-universal amplitude, $\phi$ is a phase shift, $n_0$ is the background density and $k_F$ is the Fermi wavevector. An example is given in Fig.\ref{cor-ent-tp-0.25}(a) for $L_x=60$ cylinder at doping concentration $\delta=10\%$ with $U=12$ and $t'=-0.25$. 

Due to the presence of CDW modulations, the SC correlations $\Phi_{\alpha\beta}(r)$ exhibit similar spatial oscillations with $n(x)$. Following the procedure in Ref.\cite{Jiang2018hub,Jiang2018tj}, we find that the SC correlations in this phase always decay with a power-law, whose exponent $K_{sc}$, shown in Fig.\ref{K-tp-0.25} for a range of hole doping concentrations $\delta=8.33\% \sim 12.5\%$, is obtained by fitting the results using Eq.(\ref{Eq:SC}). An example of the SC correlations at hole doping concentration $\delta=10\%$ is given in Fig.\ref{cor-ent-tp-0.25}(c). The smoking gun test for LE is whether the product of  the exponents $K_c K_{sc}=1$:  we extract  the exponents $K_c$ and $K_{sc}$ at various doping concentrations $\delta$ on cylinders of length up to $L_x=72$, and we find that the relation $K_c K_{sc}\sim 1$ is satisfied within the numerical uncertainty (see Fig.\ref{K-tp-0.25}). 

As a further test,  a key feature of the LE liquid is that it has a single gapless charge mode with central charge $c=1$, which can be obtained by calculating the von Neumann entanglement entropy $S={\rm Tr}\rho {\rm ln} \rho$, where $\rho$ is the reduced density matrix of a subsystem with length $x$. For critical systems in 1+1 dimensions described by the conformal field theory (CFT), it has been established\cite{Calabrese2004,Fagotti2011} that $S(x)=\frac{c}{6}{\rm ln}(x)+\tilde{c}$ for open systems, where $c$ is the central charge of the CFT and $\tilde{c}$ is a model dependent constant. For finite cylinders of length $L_x$, we can fix $x=L_x/2$ to extract $c$. Examples are shown in Fig.\ref{cor-ent-tp-0.25}(d) for $U=12$ and $t'=-0.25$ at different hole doping concentrations $\delta$. The extracted central charge $c\sim 1$ for all cases, as shown in the inset of Fig.(\ref{cor-ent-tp-0.25})(d), is consistent with one gapless charge mode. 

This establishes the unique signature of the LE phase in so far the charge properties are concerned. However, it should also exhibit gaps in the single fermion- and spin response. We find that the equal time spin-spin correlator in Eq.\ref{Eq:SS} decays exponentially with a finite correlation length $\xi_s\sim 8.9$ for $U=12$, $t'=-0.25$ and $\delta=12.5\%$.
%is characterized by a finite correlation length that it significantly smaller than the one found in the FS phase {\bf pls add/discuss numbers + figure}, see Fig.\ref{cor-ent-tp-0.25}. 
In addition, we do observe that the modulation of the spin density is consistent with the
Luttinger liquid/stripe rule that the spin ordering wavelength %vector 
is twice that of the charge, $\lambda_s = 2 \lambda_c$.

Finally,  the single fermion correlator Eq.(\ref{Eq:SP}) exhibits also a finite correlation length. We find that $G_\sigma(r)\sim e^{-r/\xi_G}$ with yet again a small correlation length $\xi_G$ 
as for the filled stripes. For $U=12$, $t'=-0.25$ and $\delta=12.5\%$ we find it to be $\xi_G\sim 4.8$, as compared to a spin correlation length  $ \xi_s\sim 8.9$. The charge is now 
massless and the single fermion gap is bounded from below by the spin gap, $\xi_G / 2 \le \xi_s$. However, we find that the single particle correlation length is substantially smaller
than $2 \xi_s$ This may be interpreted as the single particle 
correlation length reflecting a pairing gap while the spin gap is substantially smaller reflecting the fact that the spinons may bind into a triplet excitation inside the SC state, see the next section. Anticipating the discussion of the $LE2$ phase,  the most striking difference between the two Luther-Emery phases is in the behaviour of the single particle sector. Its large single particle gap leaves no doubt that the $LE1$ is a strongly coupled affair, reminiscent of a local pair superconductor/CDW affair. Quite different from the $LE2$ phase, we have not managed to identify the oscillations seen in $G(r)$ with any natural length scale indicating that this is as in the FS phase associated with complicated short distance physics.

In fact, only in one regard this LE1 is different from the canonical 1D textbook version. The periodicity of the CDW in the latter case is again determined by continuation from 
the free fermion limit: it should correspond with $4k_F$ and we argue that this should count as filled stripes. However, the periodicitiy of the LE1 phase counts as {\em half-filled stripes}: the  ordering wavevector is $Q_c=4\pi\delta$ of wavelength $\lambda_c=1/2\delta$ according  to Fig.\ref{cor-ent-tp-0.25}, indicating  only half a hole per each CDW unit cell!

On the cylinder ``2D-like'' transversal space for motion is possible at short distances which is apparently quite sensitive to $t'$. This short distance ladder physics should be viewed as a strongly coupled affair where many body entanglement plays a crucial role given the fact that the LE behaviour at long distances requires a very large bond dimension. These observations suggest that at short distances half-filled spin stripes are formed. We confirm the early finding that the short range SC correlations are much stronger along the rungs than in the direction of the legs.\cite{Dodaro2017,Jiang2018hub,Jiang2018tj} As was pointed out very early,\cite{White1997,Scalapino2012} it may well be that some kind of RVB like state is realized on the half-filled stripes where spin singlets are exchanging with pairs, offering a potential explanation for the preferred on-stripe charge density of 1/2 charge per domain wall unit cell. At long distances this ``fluctuating order'' may then renormalize into the universal 1D LE physics, leaving behind as a finger print that the doubled charge- and spin periodicities are  disconnected from the generic $2k_F$, $4k_F$ wavevectors of truly 1D physics. Notice that in Ref. \cite{Dodaro2017} it was pointed out that is still consistent with the so-called generalized Luttinger theorem.

%==Fig.6==
\begin{figure}[tb]
\centering
\includegraphics[width=\linewidth]{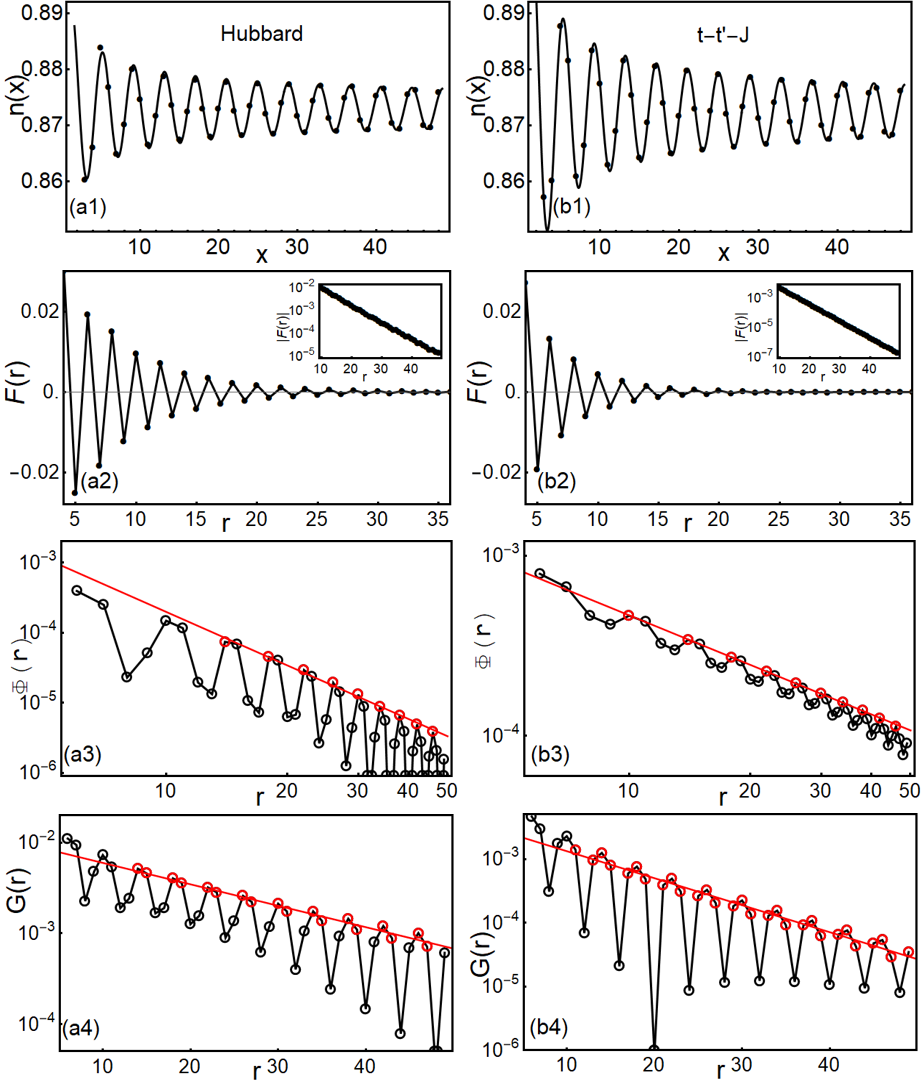}
\caption{(Color online) (a1-a4) The charge density distribution $n(x)$, spin-spin correlation $F(r)$, pair field correlation $\Phi(r)$ and single particle correlation $G(r)$ of the Hubbard model on $L_x=96$ cylinder with $\delta=12.5\%$. Here $t'=0.25$ and $U=12$. (b1-b4) The charge density distribution $n(x)$, spin-spin correlation $F(r)$, pair field correlation $\Phi(r)$ and single particle correlation $G(r)$ of the $t-t'-J$ model on the same lattice with $\delta=12.5\%$. Here $t'=0.18$ and $J=1/3$.}
\label{cor-tp0.25}
\end{figure}

\section{The second Luther-Emery phase (LE2) versus d-wave superconductivity}\label{Sec:LE2}

For positive $t'$ we find yet another phase to be very robust. As the LE1 phase it extends in a large regime of doping and $U$'s and persists up to  large $t'$ as well. As we will now show, the charge properties show precisely the diagnostics of the LE phase. These are actually of the same kind as found in the LE1 phase:  power-law SC pair-field correlations concomitant with the dual half-filled stripe algebraic CDW order, see Fig.\ref{cor-tp0.25}. We already highlighted the similarity of the phase diagrams of the Hubbard and $t-t'-J$ model \cite{Dodaro2017}.  The  charge physics of the LE2 phases behave in both models in a quite similar fashion. 
We compare these in Fig.\ref{cor-tp0.25} for representative parameters: $J=1/3$ and $t'=0.18$ for the $t-t'-J$ model, while $U=12$, $t'=-0.25$ for the Hubbard model. 
%{\bf but what do you take for Hubbard?}) 
It clearly seen that both models have similar charge- and spin-density-wave oscillations along with exponentially decaying spin correlations. Moreover, both the SC pair-field $\Phi(r)$ and charge correlations decay algebraically with corresponding exponents, e.g., $K_{sc}\approx 0.96$, $K_c\approx 1.14$ and $K_c K_{sc}\approx 1.09\sim 1$ for the $t-t'-J$ model.
%{\bf for $t-J$, Hubbard? This is a bit chaotic, please clean out.}  
The extracted central charge $c\sim 1$ as shown in SM, which further confirms the presence of the LE2 phase.

This confirms that the LE2 phase is indeed of the LE kind. However, both the single electron- and the spin correlations appear at first sight to violate this assignment. As we argued, the single fermion correlations should be short ranged as is the case in the LE1 phase revealing that Cooper pairs are formed. Taking representative parameters $U=12$, $t'=0.25$ and $\delta=12.5\%$, we find that within our resolution the single particle gap is vanishing: the correlation length $\xi_G > 33$ for a cylinder of length $L_x=96$ as shown in Fig.\ref{cor-tp0.25}. (see SM for more details) This is actually dependent on models. We find that in the $t-t'-J$ model  the single particle gap is finite but rather small $\xi_G \sim 11$ compared to the LE1 case.

Turning to the spin correlations it becomes even more of a puzzle. A most dramatic difference with the LE1 phase is that the spin periodicity observed through the equal time spin correlator at short distances is no longer set by the generic $\lambda_s=2\lambda_c$, but instead these reveal a simple two sublattice staggered antiferromagnetism, with a wavelength  $\lambda_s=2$, as shown in Fig.\ref{cor-tp0.25}. This is independent of the details of the charge stripe periodicity, while we observe it throughout the LE2 phase independent of the parameters.  

The next difficulty is that generically the spin correlation length is shorter than the single electron correlation length. In the Hubbard case we find for instance that $\xi_s\sim 5.9$
for parameters $U=12$, $t'=0.25$ and $\delta=12.5\%$ where $\xi_G>33$ within the precision of DMRG. This same pattern repeats in the $t-t'-J$ model where we find  $\xi_s\sim 3.8$ while $\xi_G\sim 10$ for parameters $J=1/3$ and $t'=0.18$. This appears to be at first sight completely unreasonable: the single particle sets a lower bound for the energy it costs to insert a spinon. The triplet gap should be at the least twice the spinon energy: it can be less because spinons may bind together in a triplet excitation, but it cannot be larger. For this simple reason it is by principle to 
find a single electron correlation length that is (much) larger than the spin correlation length, as mentioned above.
%{\bf Do not hesitate to dress this up with more simulation results.}

How to understand these results? There is actually a natural explanation: in strictly 1D all singlet superconductors are effectively $s$-wave since Cooper pairs cannot have angular momentum. This $s$-wave nature is therefore automatically hardwired into the LE state. On the ladders there is however room for angular momentum, and as we already alluded to the short-range SC correlations are clearly of the $d$-wave type. The sign of the order parameter flips sign upon going from the $x$ to the $y$ direction. How can this influence the long wavelength physics?

Dealing with a relatively weakly interacting system one can follow the canonical procedure for chain systems. One departs from the kinetic term of the free system, which 
is bosonized and then combined with the interaction terms  that take a simple form in the bosonized representation, to solve subsequently in the interacting bosonic field
theory. The free theory then leaves its imprint on the interacting system in the form of quantities like the $2k_F$ and $4k_F$ ordering wavevectors as we already stressed
in the above. Although it is not the standard procedure, one could view the LE state in a similar way as a descendent of the BCS superconductor taken as the 
free limit. It will fall short explaining why the CDW and the superconductor are in a dual relation but it does explain why LE is characterized by a spin- and single particle gap. 

%==Fig.7==
\begin{figure}[tb]
\centering
\includegraphics[width=\linewidth]{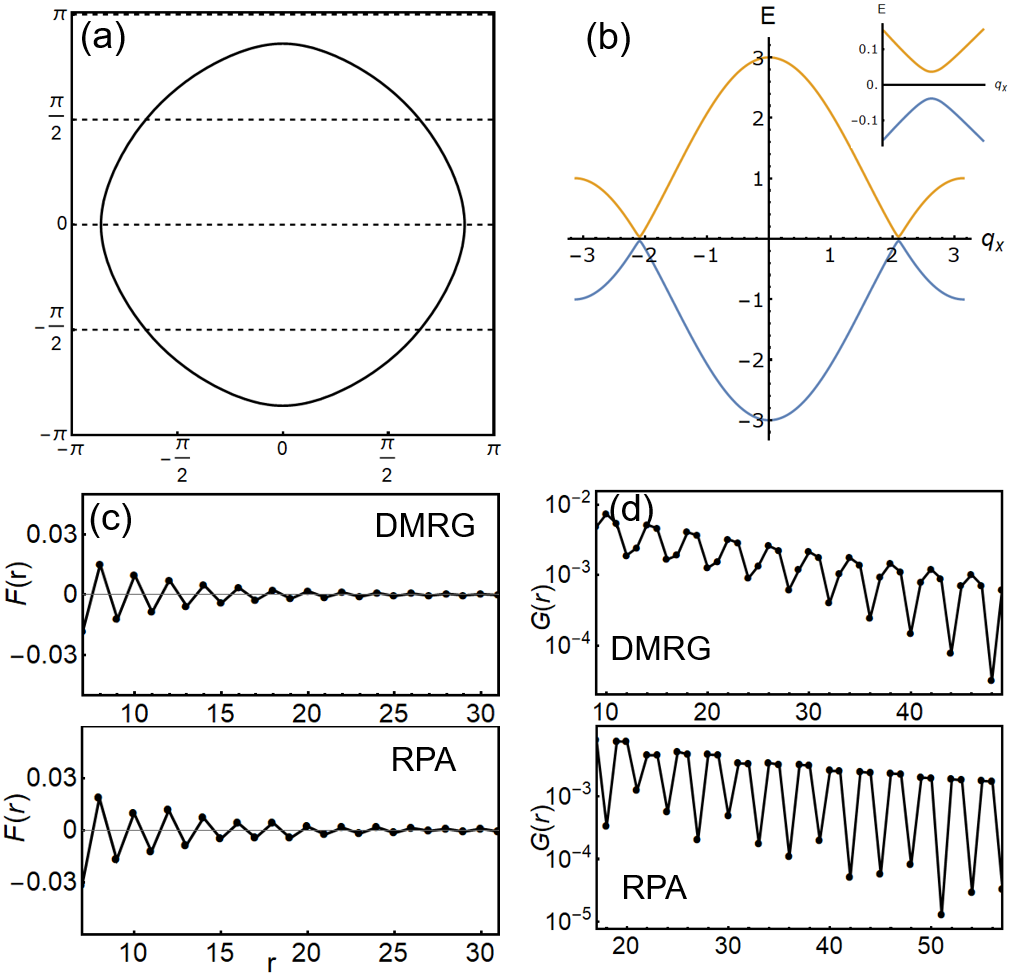}
\caption{(Color online) (a) The Fermi surface of $t=1$ and $t'=0.3$ model (details in SM). Dashed line: four quantized momenta $q_y=0,\pm \pi/2,\pi$ in the $y$ direction. (b) The quasiparticle dispersion of the $q_y=\pi/2$ mode. Inset is the zoomed-in illustration of the nearly vanishing gap of the dispersion. (c) Upper: the spin-spin correlation $F(r)$ of the Hubbard model on $L_x=96$ cylinder with $\delta=12.5\%$, same as Fig.\ref{cor-tp0.25}(a2). Lower: The spin-spin correlation obtained from RPA calculation for the 4-leg model shown in (a). (d) Upper: the single particle correlation $G(r)$ of the Hubbard model, same as Fig.\ref{cor-tp0.25}(a4). Lower: The single particle correlation $G(r)$ obtained from non-interacting model shown in (a).}
\label{RPA_spin}
\end{figure}

Let us consider how this works departing from a $d$-wave BCS state on the ladder. The amplitude of the pair density appears to be quite different along the $x$ and $y$ directions according to the DMRG simulations given the fact that the four-fold rotational symmetry of the square lattice is of course broken on the cylinder.\cite{Dodaro2017} Let us however for the sake of the argument assume that an isotropic d-wave superconductor is formed. The fermion spectrum will then be characterized by nodal lines along the diagonals of the zone. Near half-filling one then expects a nodal point on the Fermi surface in the close proximity of $ ( q_x, q_y ) = ( \pm \pi/2, \pm \pi/2 )$ on the square lattice as illustrated in Fig.\ref{RPA_spin}(a). 

On the 4-leg cylinder there are four finite-size quantized momenta available in the $y$ direction $q_y = 0, \pm \pi/2, \pi$: two of the four 1d modes in the $x$ direction characterized by $q_y = \pm \pi/2$ will become massless at $q_x = \pm \pi/2$ because it intersects the nodal points! (see Fig.\ref{RPA_spin}(b)) The system can of course be still a superconductor since it is gapped at the other allowed momenta. The equal time fermion propagator will now show an infinite correlation length at large distances since it is completely dominated by the presence of the massless points, and this should be remembered by the bosonized interacting theory. This vanishing mass is not robust. Given that there is no $C_4$ symmetry axis on the cylinder, there has to be generically a s-wave admixture giving rise to a finite mass at the nodal point, while also this minimum gap may shift away from the nodal line. But this gap will be generically quite a bit smaller than the average $d$-wave gap. This explains the observation of the single particle correlation length that may seem to diverge, or stay quite finite pending the details of the model.
This interpretation is directly confirmed by the oscillation observed in $G_\sigma(x)$. The massles ``nodal point'' should be in the close vicinity of $q_x =\pm \pi/2$, implying a wavelength $\lambda_G \simeq 4$. This is precisely what is observed and the DMRG result for $G_\sigma(r)$ is very similar as the result for the non-interacting d-wave model as shown in Fig.\ref{RPA_spin}(d). %{\bf We should show these side to side in a figure in the main text, as for the spin correlations.  Perhaps put it all together in a separate figure?}  

%{\bf Do we dare to make a direct comparision between this d-wave equal time correlator and the DMRG fermions? In a perfect world they would look quite similar, the only difference being that the exponent of the algebraic decay would be different.} [{\textbf{HCJ: Jan, I think the explanation you provided here makes a lot sense already, maybe we don't have to provide such a direction comparison, we can do later if the referees ask for it.}}]

How to explain the spin correlations? It is an equally well established wisdom that the collective response of a not too strongly interacting 1D system are quite well approximated
by the results of the RPA. One computes first the dynamical susceptibility of the free system $\chi_0 (\vec{q}, \omega)$ (the Lindhardt function)
and the effects of the interactions are included on the time dependent mean field level $\chi (\vec{q}, \omega) = \chi_0 (\vec{q}, \omega) / ( 1 - J_{\vec q} \chi_0 (\vec{q},\omega) )$
with the interaction parameter $J_{\vec q}$ pending the nature of the response and interactions. The way this works in the $d$-wave SC state on the square 
lattice is well known since it has turned into the standard explanation for the so-called magnetic resonance observed in cuprate superconductors.\cite{Bulut1996,Brinckmann1999} A gap opens up in the Lindhardt
spectrum associated with the SC gap in the single particle momentum. This is at maximum at $\vec{q} = (\pi, \pi)$, the momentum associated with a simple staggered order parameter. In the proximity of momenta like $(\pi , 0)$ it will close given the fact that node-to-node scattering is kinematically allowed. 

A next ingredient is that upon switching on the enhancement factor $J_{\vec q}$ a bound mode appears in the Lindhardt gap that is generically most strongly bound at the $(\pi, \pi)$ point for the reason that the density of electron-hole pair states is highest at these momenta. Similarly, the $\chi''$ is strongly suppressed at the massless points for simple kinematical reasons. We computed this RPA susceptibility for the cylinder system using the electron dispersion relations we just discussed, finding that with regard to these features it is very similar as on the square lattice. Practically, the simple staggered magnetic order becomes to be dominant once $J_{(\pi,\pi)}$ exceeds $2.78t$. (see SM for more details) 
%{\bf We should add  $\chi_0", \chi"$ false colour plots to the SM, here in the main text we should at the least specify how much enhancement is necessary to get the simple AFM outcomes.} 

What does this mean for the equal time spin-spin correlations? We computed the spin-spin correlation functions from the RPA dynamical susceptibility, finding for a reasonable choice of the interaction parameter $J_{\vec q }$ an outcome that is virtually identical to the DMRG result in the LE2 phase, see Fig.\ref{RPA_spin}(b). The way this works is simple. One anticipates that the massless spin excitations should dominate the spin correlator at large distances. However, for the reason that the associated density of states is strongly suppressed this contribution is just not resolved. Instead, given that the dynamical susceptibility is dominated by the ``resonance'' at the staggered ordering wave vector it just dominates also the equal time correlations. 

The take home message is that the DMRG results become comprehensible asserting that at short distances the system renormalizes into a seemingly rather weakly interacting $d$-wave superconductor living on the ladder. The main effect of the residual interactions appears to be in the spin channel where enhancement factors take care that a well developed ``magnetic resonance'' forms. However, this is not yet the whole story: this ``nearly non-interacting'' d-wave Luther-Emery view is in one regard quite misleading. The LE2 phase is still characterized by the ``half-filled stripe'' algebraic CDW order. As for the spin system one could look for a nesting type fermiology interpretation but this turns out to fail completely.\cite{Dodaro2017}  There is just nothing in the non-interacting d-wave band structure that relates to the ordering momentum of the charge density. In this regard the LE2 phase continues to be rooted in strong coupling physics.

\section{Conclusions}

Resting on the fact that it has become recently possible to compute reliably the ground states of 4-leg ladder Hubbard systems by DMRG we have systematically investigated the 
ground state properties in the low doping regime. Given that these ladder systems renormalize into one dimensional systems at large distances one can rest of the well understood universal properties of 1D physics to diagnose the physics.  At short distances the physics is like that of the strongly interacting  2D system that is apparently ruled by dense many body entanglement as signalled by the need for very large bond dimensions in the DMRG. The outcome is in the form of LE liquids and the ``commensurately pinned Luttinger liquid'' (the filled stripes) that obey the 1D universality although these rest on short distance data which are entirely different from the canonical chain systems. 

Surprisingly, we find that the zero temperature phase diagram is rather insensitive to the strength of the interactions when these large enough, and the doping when it is not too high. Instead, the ground states turn out to be exceedingly sensitive to the next-nearest-neighbour hopping $t'$. We do not find a plethora of phases as sometimes be claimed: as function of $t'$ there are just three preferred phases. Why this is the case is not at all understood:  standard product state mean field (Hartree-Fock) does not give any clue in this regard and it has to be somehow rooted in the densely entangled nature of the short distance physics.

One part of the puzzle is why the simple motive of commensurate pinning stabilizing the filled stripes is so extremely sensitive to $t'$.  When this looses out, the liquid phases that are formed are of a very different nature pending the sign of $t'$. The ``LE1 liquid'' for negative $t'$ shows the symptoms of a strongly interacting, yet in a way conventional LE state. The behaviour of correlations in the LE liquid suggest a highly collective fluctuating order physics: well developed half-filled spin stripes which are formed from Cooper pairs (the Luther-Emery charge density -- pair correlation function duality), kept fluctuating by 1D zero point motions. 

Strikingly, the ``LE2 liquid'' setting in at small positive $t'$ appears to be instead a continuation of a weakly interacting $d$-wave superconductor living on the ladder. The symptoms are in the first place the observation of a small (or even vanishing) gap in the single particle correlations, suggesting that the system remembers the nodal fermions of the 2D BCS $d$-wave superconductor. The spin correlations are even more informative: the two sublattice spin correlations are seemingly uniquely explained by the generic fermiology motive of strong enhancement of staggered spin correlations in the $d$-wave SC state.

This brings us to experiment. One may wonder whether the ladder is no more than a metaphor or whether there may be a more literal relation with the observations. In the mid 1990's the half filled spin stripes were discovered in the 214 superconductors. It is well understood that the incommensurate static spin order in these systems gives in precisely to the stripe-rule that $\lambda_s = 2 \lambda_c$ while the spin order parameter is quite large after any standard. Moreover, the doping dependence of the periodicity at least for $\delta \le 12.5\%$ is linear indicating a preferred density of half a hole  per domain wall unit cell. This was also the origin of the idea of fluctuating order: the spin excitations measured by inelastic neutron scattering are suggestive that also in the absence of static order there are strong stripe correlations going on at higher energy. 

However, in the 123 and 2212 families especially near optimal doping the case evolved that one may be  closer to the truth using the fermiology language.  The spin excitations are dominated by the magnetic resonance occurring at the staggered wave vector which near optimal doping is only present in the SC state. This is naturally explained by the same RPA logic as we used to explain the magnetic correlations in LE2. 

There has been much debate in the past trying to reconcile these seemingly very different forms of physics of these two families of cuprates that seem chemically so similar: all the action is supposed to take place in the same $Cu-O$ layers. However, we now learn from the Hubbard ladders that tiny differences in a microscopic 
parameter that based on established wisdom should not matter can make a big difference, apparently causing the system to land in near opposite regimes of long distance 
physics.

{\it{Note added.}} After completion of this work, we became aware of a work by Ponsioen et al.\cite{Ponsioen2019}, who studied the same model on the 2D square lattice using iPEPS and proposed that the ground state of system at doping $\delta\leq 0.14$ is in a state with period-4 charge stripe but without SC for negative $t'$, while in a state with coexisting uniform SC with long-range antiferromagnetic order but without charge stripe for positive $t'$. This is different with this work in both negative and positive $t'$ sides, where we find coexisting quasi-long-range SC and CDW correlations but without (quasi-)long-range spin order.

{\it Acknowledgments:} We thank Steven Kivelson, Douglas Scalapino, and Zheng-Yu Weng for insightful discussion and suggestion. This work was supported by the Department of Energy, Office of Science, Basic Energy Sciences, Materials Sciences and Engineering Division under Contract DE-AC02-76SF00515. Parts of the computing for this project was performed on the Sherlock cluster.

%%Reference
\bibliography{Refh4c}

\newpage

\renewcommand{\theequation}{A\arabic{equation}}
\setcounter{equation}{0}
\renewcommand{\thefigure}{S\arabic{figure}}
\setcounter{figure}{0}
\renewcommand{\thetable}{A\arabic{table}}
\setcounter{table}{0}
\section{Supplemental Material}

%%==Fig.S1==
\begin{figure}[b]
\centering
\includegraphics[width=\linewidth]{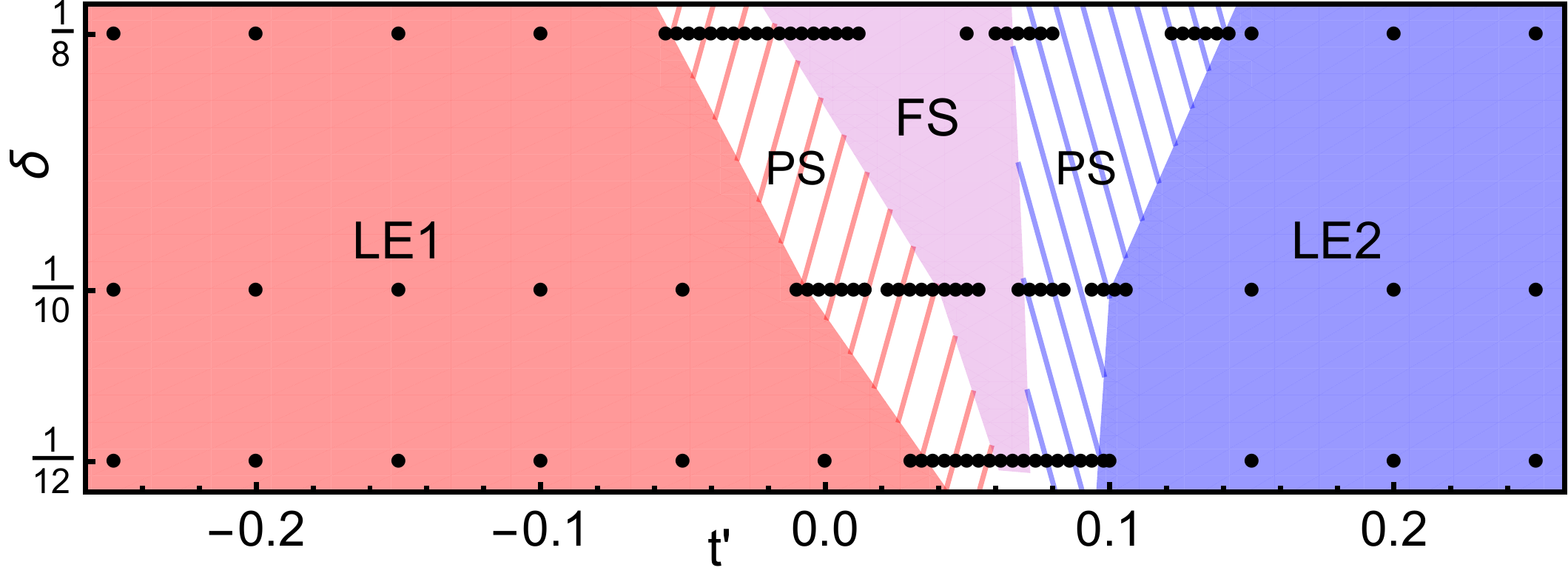}
\caption{Ground state phase diagram of Hubbard model at $U=12$ as a function of hole doping concentration $\delta$ and $t'$. The black dots are data points.}
\label{SM-phaseU8}
\end{figure}

\subsection{Numerical details}
For most of the DMRG calculations we keep $m=10000\sim 20000$ number of $U(1)$ states and perform around 60 sweeps to obtain the ground-states. As the ground-state of finite system should not spontaneously break symmetry of the Hamiltonian, such as $SU(2)$ spin rotational symmetry and translational symmetry around the cylinder, we should expect a zero magnetic moment $\avg{S_i^z}=0$ etc. This is indeed the case in our simulation when we keep $m\geq 10000$ number of states in both LE phases, which allows us to obtain reliable results including the correlation functions.

On the contrary, full convergence in the ``filled'' stripe phase is known to be more challenging partially due to the enlarged charge density oscillation periodicity and the enhanced quantum fluctuation near the phase boundaries as the filled stripe phase is relatively small in the phase diagram. To resolve this issue, we further implement higher $SU(2)$ symmetry in the DMRG simulation and push the effective $U(1)$ number of states up to $m\sim 36000$, which gives excellent convergence in our study on system such as $L_x=96$ cylinder.

\subsection{Phase diagram of the Hubbard model at $U=8$}\label{SM:U8Phasediagram}

The ground state phase diagram of the Hubbard model at $U=8$ is shown in Fig.\ref{SM-phaseU8} for three different hole doping concentrations $\delta=8.33\% \sim 12.5\%$ on 4-leg cylinder of length $L_x=48$, 40 and 32, respectively, where the phase boundaries are determined by the distinct pattern of charge density modulation of the adjacent phases. This is quite similar with the phase diagram in Fig.\ref{Fig:PhaseDiagram} of $U=12$ shown in the main text, but with a slightly larger filled stripe phase.

\subsection{The filled stripe phase}\label{SM:Filledstripe}

%%==Fig.S2==
\begin{figure}[tb]
\centering
\includegraphics[width=\linewidth]{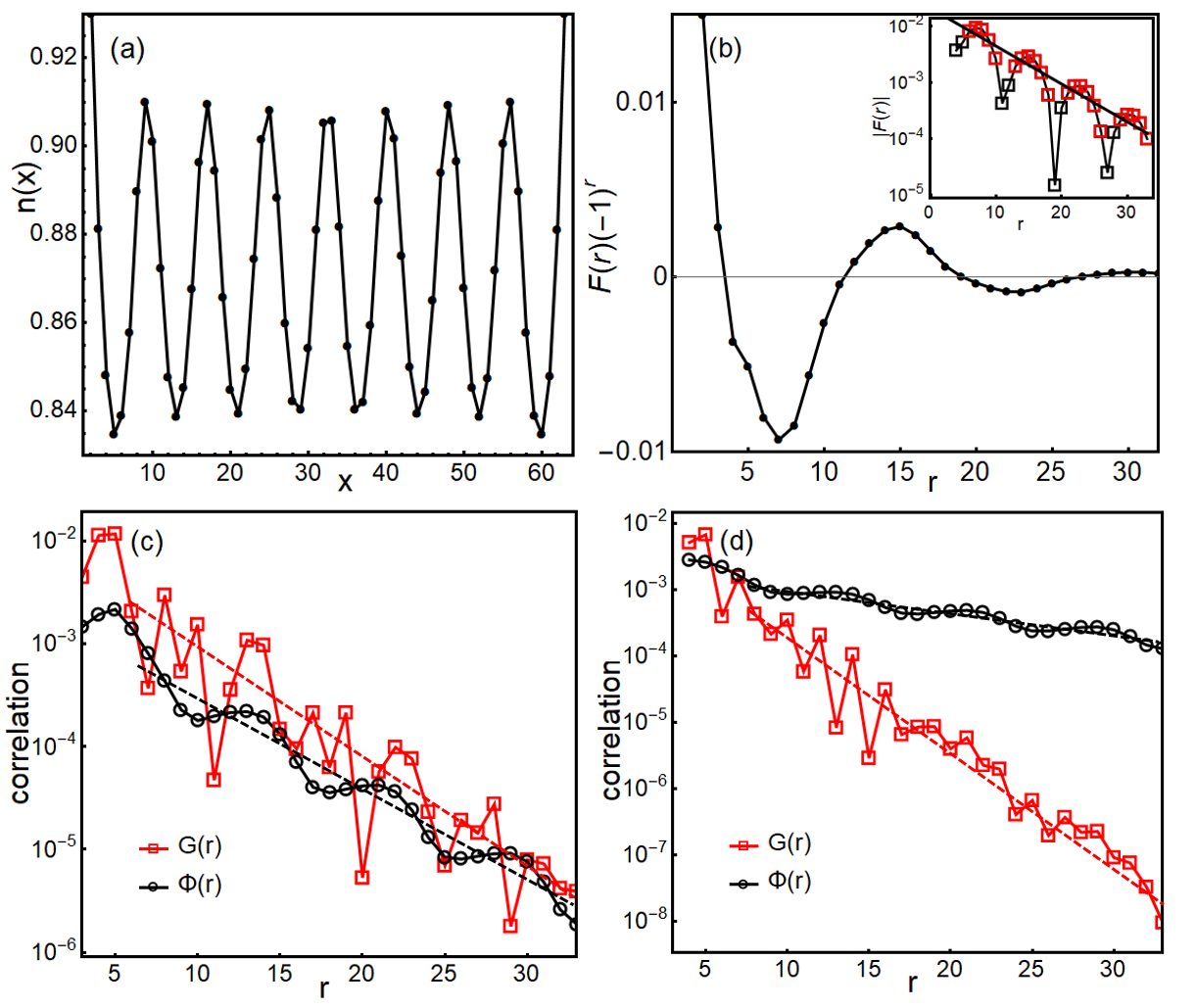}
\caption{(Color online) (a) Charge density profile $n(x)$, (b) spin-spin correlation $F(r)$ and (c) superconducting pair-field $\Phi_{yy}(r)$ (black) and single particle $G(r)$ (red) correlations of the Hubbard model in the filled stripe phase on a $L_x=64$ cylinder at $U=12$, $t'=0.05$ and $\delta=12.5\%$ by keeping $m=18000$ number of states. (d) Superconducting pair-field $\Phi_{yy}(r)$ and single-particle $G(r)$ correlations of the $t-t'-J$ model at $J=1/3$, $t'=0.07$ and $\delta=12.5\%$. Dashed lines are guides for eyes.}
\label{SM-tp0.05}
\end{figure}

As mentioned in the main text, it is more challenging to obtain fully converged ground state of the Hubbard model in the filled stripe phase than the LE phases and especially directly establishing the long-distance behavior of correlation functions, such as the superconducting correlation. To resolve this issue and directly nail down the nature of the filled stripe phase, we focus on the deep of this phase with a characteristic set of parameters $t'=0.05$, $U=12$ and $\delta=12.5\%$. By keeping around $m=18000$ number of states, we are able to converge to the true ground state which preserves all the symmetries of the Hamiltonian including the spin $SU(2)$ rotational symmetry with $\langle S_i^z\rangle =0$. Examples are shown in Fig.\ref{SM-tp0.05} for the system on a $L_x=64$ cylinder. Consistent with the filled stripe phase,\cite{Zheng2017,Jiang2018hub} the charge density profile shows a modulation of wavelength $\lambda_c=1/\delta$, and the spin-spin correlation decays exponentially $F(r)\sim e^{r/\xi_s}$ with a finite correlation length $\xi_s\sim 6.5$. Both the single-particle and superconducting pair-field correlations are also short-ranged: $G(r) \sim e^{-r/\xi_G}$ and $\Phi(r) \sim e^{-r/\xi_{sc}}$ with finite correlation length $\xi_G \sim 4.0$ and $\xi_{sc} \sim 4.6$, respectively. Again, they are consistent with the filled stripe phase in the $t-t'-J$ model discussed in the main text.

\begin{figure}[tb]
\centering
\includegraphics[width=\linewidth]{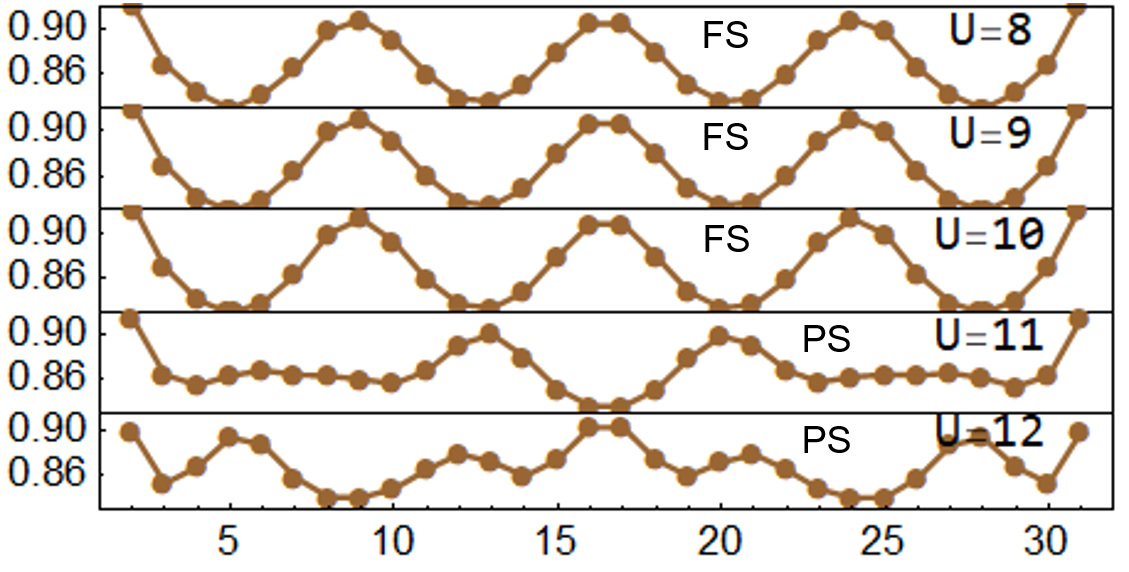}
\caption{(Color online) Charge density profiles $n(x)$ of the Hubbard model at $t'=0$ and $\delta=12.5\%$ for $U=8\sim 12$. The filled stripe is no longer stable when $U>10$.}
\label{SM-filled-n-U}
\end{figure}

However, this filled stripe phase is only stable in a small region around or close to $t'=0$, which becomes increasingly fragile with the increase of $U$ or the decease of the hole doping concentration. For example, the ground state of the system at $t'=0$, which is in the filled stripe phase at $U=8$, is no longer the case when $U>10$. Even at $U=8$, a tiny $t'\sim -0.01$ is large enough to drive the system out of the filled stripe phase. Examples of charge density profile $n(x)$ of the Hubbard model on a $L_x=32$ cylinder are given in Fig.\ref{SM-filled-n-U} with $\delta=12.5\%$ and $t'=0$ at different $U$. For each simulation, we start with a state with filled charge stripe pinned by appropriate chemical potential, and then check the stability of the filled stripe in the following DMRG sweeps by turning off the pinning field. Our results show that the filled stripe disappears when $U>U_c\sim 10$.

%%==Fig.S3==

%%==Fig.S4==
\begin{figure}[tb]
\centering
\includegraphics[width=\linewidth]{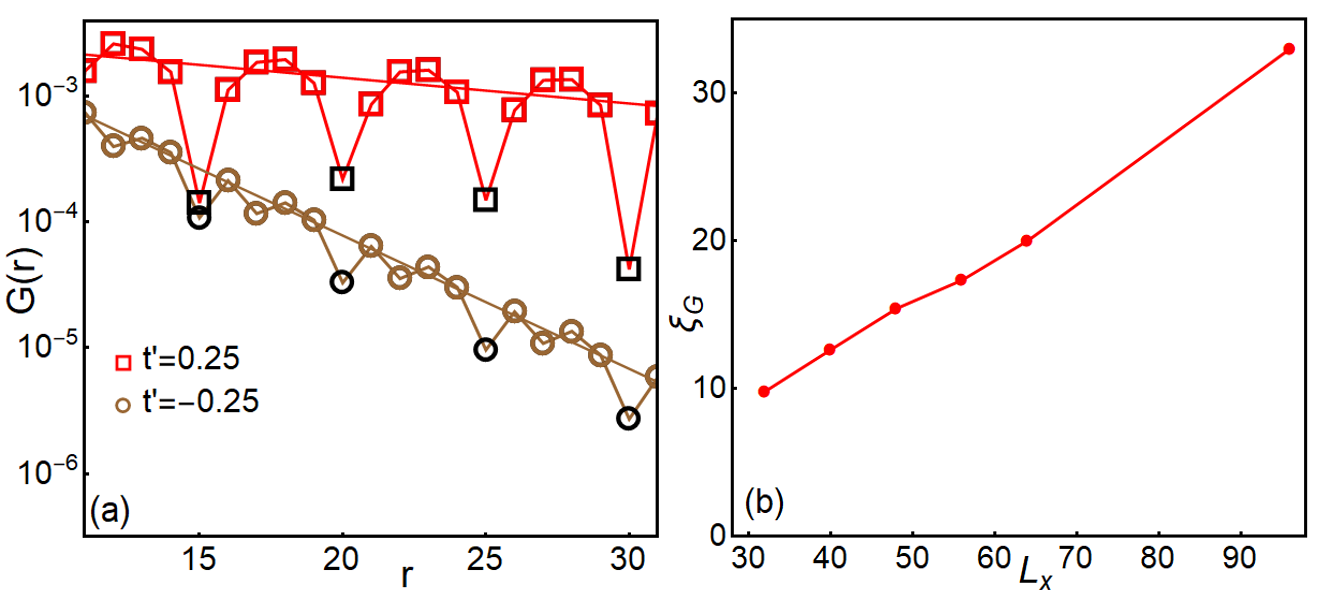}
\caption{(Color online) (a) Single-particle correlations $G(r)$ of the Hubbard model on a $L_x=60$ cylinder at $U=12$ and $\delta=10\%$ in the LE1 ($t'=-0.25$) and LE2 ($t'=0.25$) phases, respectively. (b) Single-particle correlation length $\xi_G$ as a function of cylinder length $L_x=32\sim 96$. The Hubbard model is fixed at $t'=0.25$, $U=12$ and $\delta=12.5\%$ point.}
\label{SM-cor}
\end{figure}

\subsection{Single-particle correlation}

One puzzling point is the slowly decaying single-particle correlation of the Hubbard model in the LE2 phase. Fig.\ref{SM-cor} (a) shows $G(r)$ at $t'=-0.25$ in the LE2 phase with $U=10$ and $\delta=10\%$ on a $L_x=60$ cylinder. For comparison, the single-particle correlation $G(r)$ in the LE1 phase at $t'=-0.25$ is also given.  Fig.\ref{SM-cor} (b) shows the correlation length $\xi_G$ extracted by fitting the results using function $G(r)\sim e^{-r/\xi_G}$ for the LE2 phase. While a clear saturation tendency of $\xi_G$ vs $L_x$ is seen in the LE1 phase, it remains still unclear for the LE2 phase for cylinders of length up to $L_x=96$.

%Interestingly, our results clearly show that the single-particle correlation $G(r)$ of the $t-t'-J$ model decays fairly quick in both LE1 and LE2 phases with finite correlation length $\xi_G$, e.g., $\xi_G\sim 10$ in the LE2 phase. An example of the comparison between these two models is shown in Fig.\ref{SM-cor}(b) for system on a cylinder of length $L_x=96$, whose correlation length $\xi_G$ has been provided in the main text in the section of LE2 phase.

%%==Fig.S5==
\begin{figure}[tb]
\centering
\includegraphics[width=\linewidth]{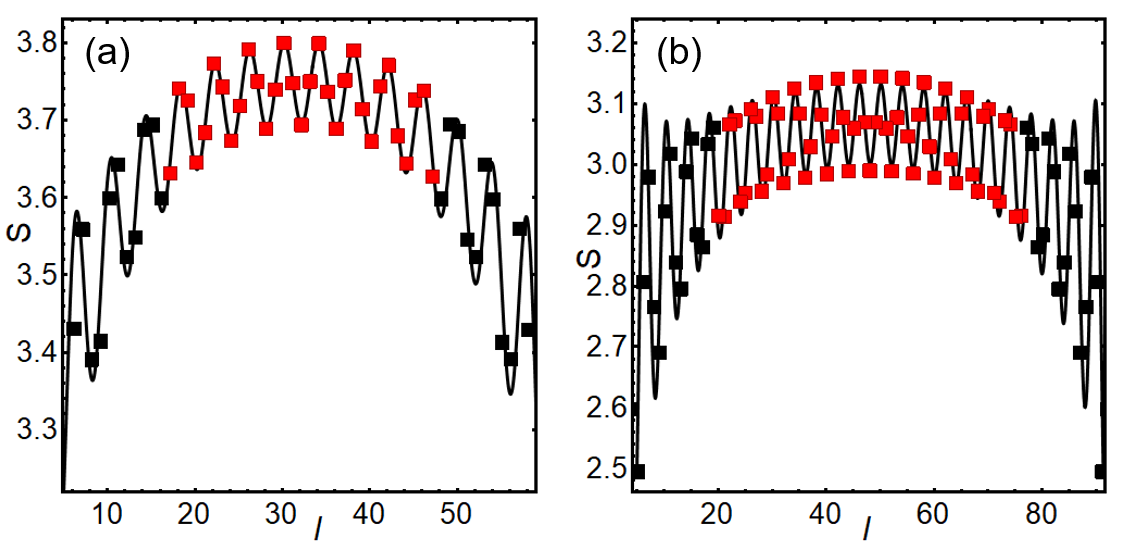}
\caption{(Color online) (a) Von Neumann entanglement entropy of the Hubbard model on a $L_x=64$ cylinder, with $t'=0.25$, $U=12$ and $\delta=12.5\%$. The central charge extracted from the red data points is $c\sim 1.8$. (b) Entropy of the $t-t'-J$ model at $J=1/3$, $t'=0.18$ and $\delta=12.5\%$ in the LE2 phase. The extracted central charge is $c\sim 1.07$. %[{\textbf{HCJ: Is the entropy of  the Hubbard model on $L_x=96$ cylinder really converged? I would suggest removing only a couple of CDW period data points in the central charge fitting to see what happens. A central charge of $c\sim 2$ is likely able to explain the (close) vanishing spin particle gap in the LE2 phase.}}]
}
\label{SM-ent}
\end{figure}

\subsection{Entanglement entropy}

As a support on our results in the main text, we have also calculated the von Neumann entanglement entropy of the Hubbard and $t-t'-J$ model in the LE2 phase as shown in Fig.\ref{SM-ent}. We choose a characteristic sets of parameter $J=1/3$, $t'=0.18$ and $\delta=12.5\%$ for the $t-t'-J$ model on the $L_x=96$ cylinder. For the Hubbard model,  to obtain fully converged entropy we choose a slightly shorter $L_x=64$  cylinder and set parameter $t'=0.25$, $U=12$ and $\delta=12.5\%$. The central charge can be extracted using\cite{Fagotti2011}
\bea
S(l)&=&\frac{c}{6} \ln \big[\frac{4(L_x+1)}{\pi} \sin \frac{\pi(2l+1)}{2(L_x+1)}|\sin k_F|\big] \nn\\
&+&b \frac{\sin[k_F(2l+1)]}{\frac{4(L_x+1)}{\pi} \sin \frac{\pi(2l+1)}{2(L_x+1)}|\sin k_F|}+ a,
\label{SM-Eq-EE}
\eea
where $l$ is the length of subsystem and $c$ is central charge. $k_F$ denotes the Fermi momentum. $a$ and $b$ are model dependent fitting parameters. The second term denotes the contribution from higher order oscillating term. The central charge $c$ extracted from $t-t'-J$ model in Fig.\ref{SM-ent}(b) is $c\sim 1.05$, which agrees with the LE liquid phase nicely. While in Hubbard model, the extracted central charge $c\sim 1.8$.

%%==Fig.S6==
\begin{figure}[tb]
\centering
\includegraphics[width=\linewidth]{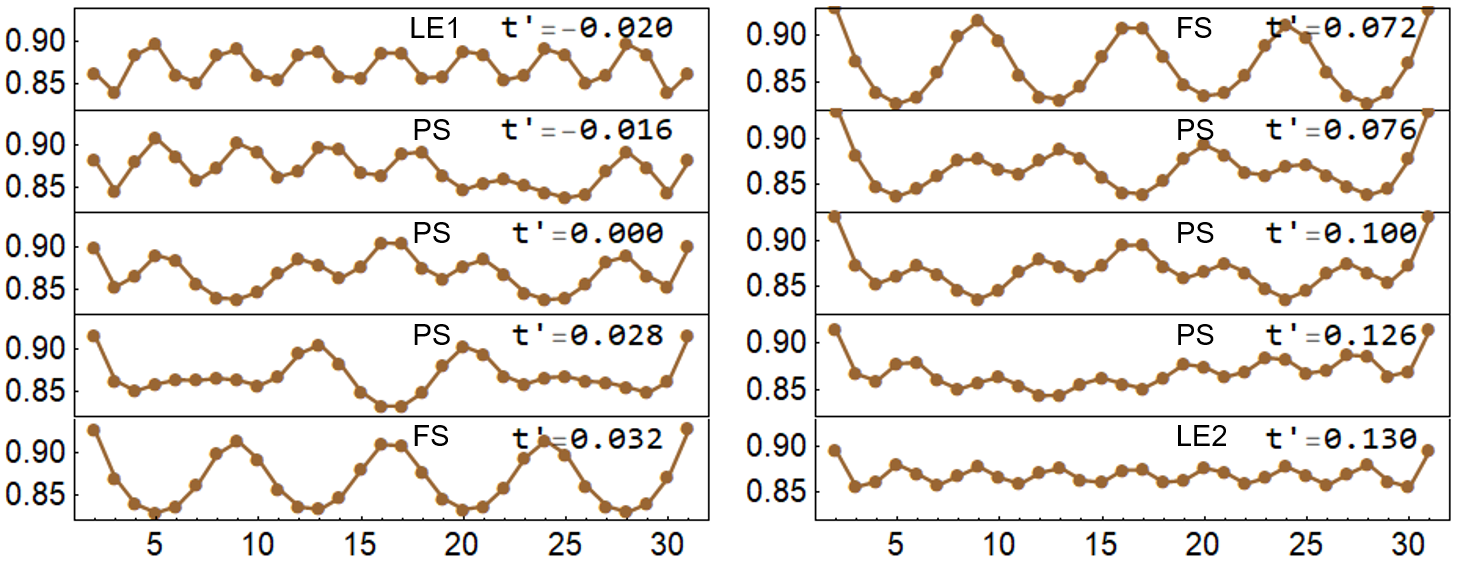}
\caption{(Color online) Charge density profiles $n(x)$ of the Hubbard model on a $L_x=32$ cylinder at $U=12$ and $\delta=12.5\%$ at different $t'$.}
\label{density_ps}
\end{figure}

\subsection{Phase separation}

In the phase diagram, there are two additional regions for phase separation sandwiched by the filled stripe and LE phases, where the evidences are provided in Fig.\ref{density_ps} for the Hubbard model. While both the half-filled charge stripe of wavelength $\lambda_c=1/2\delta$ and the filled charge stripe of wavelength $\lambda_c=1/\delta$ are clearly seen in the LE and filled stripe phases, respectively, the charge density profile $n(x)$ in the phase separation region is clearly spatially inhomogeneous, which can be considered as a combination of both filled and half-filled charge stripes. This is true for both the shaded regions in the phase diagram.

%%==Fig.S7==
\begin{figure}[tb]
\centering
\includegraphics[width=\linewidth]{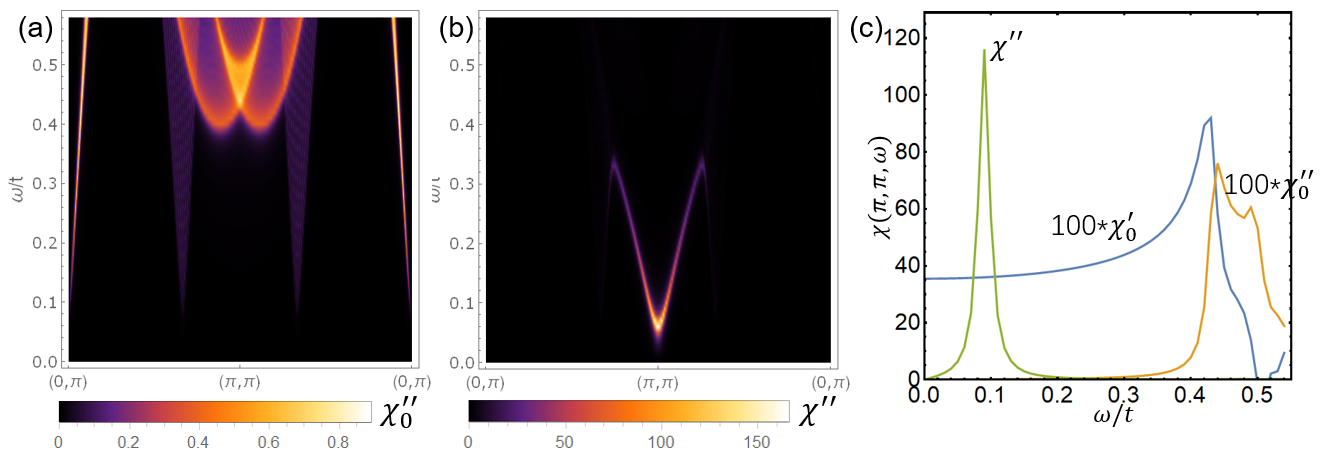}
\caption{(a) The non-interacting spin susceptibility $\chi_0''(\vec{q},\omega)$ along $(q_x, \pi)$ line. (b) The RPA spin susceptibility calculate at $J=2.78t$ point. (c) The evolution of $\chi(\pi,\pi,\omega)$ as $J$ changed from 0 to $2.78t$. }
\label{SM_RPA}
\end{figure}

\subsection{RPA calculation}
To make direct connection with DMRG results, We employ the tight-banding dispersion
\bea
\epsilon(\vec{k})&=& - \mu -2t[\cos(k_x)+\cos(k_y)]-4t'\cos(k_x)\cos(k_y), 
\eea
here we use the parameter set inside the LE2 phase, $t=1$ and $t'=0.3$. The chemical potential is $\mu= 1.0$. We use a gap function with d-wave symmetry, $\Delta(\vec{k})=\Delta_0(\cos(k_x)-\cos(k_y))/2$ with $\Delta_0=0.1$. The lattice we used is $L_x\times L_y=400\times 4$, for simplicity, we consider periodic boundary condition on both x and y direction. The quasiparticle dispersion is given by $E_{\vec{k}}=\sqrt{\epsilon^2(\vec{k})+\Delta^2(\vec{k})}$. As discussed in the main text, $k_y$ can only take four discrete values and nodel point on the Fermi surface is in the cloase proximity of $(\pi/2,\pi/2)$. The energy dispersion along the $k_y=\pi/2$ line exhibits an almost invisible tiny gap $\sim 0.03t$ located at $k_x$ slightly larger than $\pi/2$.

To analyze the position of the magnetic resonation peak and the spin-spin correlation it is instructive to start with the bare Lindhard susceptibility of the superconducting state
\bea
\chi_0(\vec{q},\omega)&=&\sum_{\vec{k}} [\frac{1}{2}(1+\Omega_{\vec{k},\vec{q}})\frac{f(E_{\vec{k}+\vec{q}})-f(E_{\vec{k}})}{\omega-(E_{\vec{k}+\vec{q}}-E_{\vec{k}})+i 0^+}\nn\\
&+&\frac{1}{4}(1-\Omega_{\vec{k},\vec{q}})\frac{1-f(E_{\vec{k}+\vec{q}})-f(E_{\vec{k}})}{\omega+(E_{\vec{k}+\vec{q}}+E_{\vec{k}})+i 0^+}\nn\\
&+&\frac{1}{4}(1-\Omega_{\vec{k},\vec{q}})\frac{f(E_{\vec{k}+\vec{q}})+f(E_{\vec{k}})-1}{\omega-(E_{\vec{k}+\vec{q}}+E_{\vec{k}})+i 0^+}]
\eea
where $f$ is the Fermi dispersion and we define $\Omega_{\vec{k},\vec{q}}=(\epsilon_{\vec{k}+\vec{q}} \epsilon_{\vec{k}}+\Delta_{\vec{k}+\vec{q}}\Delta_{\vec{k}})/(E_{\vec{k}+\vec{q}} E_{\vec{k}})$. The three parts in $\chi_0(\vec{q},\omega)$ are due to quasiparticle scattering, quasiparticle pair creation, and quasiparticle pair annihilation, respectively. To calculate the bare susceptibilty $\chi_0$ numerically, we replace $0^+$ by $\Gamma=0.01t$ in the denominators. The calculated bare susceptibility $\chi_0''(\vec{q},\omega)$ with maximum response at $\vec{q}=(\pi,\pi)$ is illustrated in \Fig{SM_RPA}(a).

In the RPA approach, the interacting spin susceptibility is given by
\bea
\chi(\vec{q},\omega)=\frac{\chi_0(\vec{q},\omega)}{1-J(\vec{q})\chi_0(\vec{q},\omega)}
\eea
where $J(\vec{q})$ is the spin-spin response function assumed to be of the form $J(\vec{q})=-J[\cos(q_x)+\cos(q_y)]/2$. After $J(\vec{q})$ is turned on, a much sharper resonance peak at $(\pi,\pi)$ is expected at low energy smaller than twice of the maximum of SC gap. In \Fig{SM_RPA}(b), we show the strong magnetic response $\chi''(\vec{q},\omega)$ calculated at $J=2.8$t point.

Finally, we calculate the equal time spin-spin correlation $F(r,t=0)$ from $\int \sum_{\vec{q}} e^{i q_x r} \chi''(\vec{q}, \omega) d\omega$. At non-interacting $J(\vec{q})=0$ point, the spin-spin correlation is generally incommensurate.  When $J(\vec{q})$ is large enough, the period-2 oscillation originated from the sharp resonant peak at $(\pi,\pi)$ becomes dominate. In our calculation, the onset of commensurate period-2 oscillation happens around $J\sim 2.78t$, which indicate a spin gap $\Delta_s \sim 0.09t$ shown in \Fig{SM_RPA}(c).

%In our phase diagram, the LE phase and the ``filled'' stripe phase are separated by a phase separation region. This is reasonable because the charge order of the two phases break translational symmetry in a different way, a first order phase transition between the two phases is generally expected. In \Fig{density_ps} we provide more evidence about the phase separation appeared in the intermediate regions.
%From the top to the bottom of \Fig{density_ps} (a), we show how the density profile changes as we continuously tune the system from LE1 phase to the ``filled'' stripe phase, i.e. the red shaded region in the $U=12$ phase diagram. The charge density profiles in this intermediate region are inhomogeneous combination of ``filled'' stripe and ``half-filled'' stripe, which are consistent with a phase separation. Similar modification of charge density profiles are observed in the shaded region sandwiched by the ``filled'' stripe and LE2 phase, as shown in \Fig{density_ps} (b).

\end{document}